\documentclass[preprints,article,accept,moreauthors,10pt,a4paper]{mdpi}

\graphicspath{{figures/}}

\usepackage{bbm}

\newcommand{\ket}[1]{\mbox{$ | #1 \rangle $}}
\newcommand{\bra}[1]{\mbox{$ \langle #1 | $}}
\newcommand{\be}{\begin{equation}}
\newcommand{\ee}{\end{equation}}

\newcommand{\PP}{\ensuremath{\mathcal{P}}}
\newcommand{\QQ}{\ensuremath{\mathcal{Q}}}
\newcommand{\CC}{\ensuremath{\mathcal{C}}}
\newcommand{\BB}{\ensuremath{\mathcal{B}}}

\usepackage{color}
\usepackage{soul}

\firstpage{1} 
\pubvolume{xx}
\issuenum{1}
\articlenumber{1}
\pubyear{2018}
\copyrightyear{2018}
\history{}

\Title{Entropic Steering Criteria: Applications to Bipartite and Tripartite Systems}

\Author{Ana C. S. Costa*, Roope Uola and Otfried G\"uhne}

\AuthorNames{Ana C. S. Costa, Roope Uola and Otfried G\"uhne}

\address[1]{%
Naturwissenschaftlich-Technische Fakult\"at, Universit\"at Siegen, 57068 Siegen, Germany; roope.uola@gmail.com~(R.U.); otfried.guehne@uni-siegen.de (O.G.)}

\corres{\hangafter=1 \hangindent=1.05em \hspace{-0.82em} Correspondence: ana.sprotte@gmail.com or ana.costa@physik.uni-siegen.de}

\abstract{The effect of quantum steering describes a possible action at a distance via local measurements. Whereas many attempts on characterizing steerability have been pursued, answering the question as to whether a given state is steerable or not remains a difficult task. Here, we investigate the applicability of a recently proposed method for building steering criteria from generalized entropic uncertainty relations. This method works for any entropy which satisfy the properties of (i)~(pseudo-) additivity for independent distributions; (ii) state independent entropic uncertainty relation (EUR); and (iii) joint convexity of a corresponding relative entropy. Our study extends the former analysis to Tsallis and R\'enyi entropies on bipartite and tripartite systems. As examples, we investigate the steerability of the three-qubit GHZ  and W states.}


\begin{document}

\section{Introduction}

The notion of steering was first introduced by Schr\"odinger in 1935 in order to capture the essence of the Einstein--Podolsky--Rosen argument~\cite{schroedingerletter}. It describes the ability of one experimenter, Alice, to~remotely affect the state of another experimenter, Bob, through local actions on her system supported by classical communication. Steering is based on a quantum correlation strictly between entanglement and non-locality, meaning that not every entangled state can be used for steering and not every steerable state violates a Bell inequality~\cite{wiseman}. 

Recently, it has been shown that steering plays a fundamental role in various quantum protocols and in entanglement theory. In the former, steering characterizes systems useful for one-sided device-independent quantum key distribution~\cite{Branciard2012}, subchannel discrimination~\cite{Piani2015} and randomness generation~\cite{Law2014}. Concerning entanglement theory, steering has been used to find counterexamples to Peres conjecture, which was an open problem for more than fifteen years~\cite{moroder, Vertesi2014, Yu2017}. Steering is also known to be closely related to incompatibility of quantum measurements. Namely, any set of non-jointly measurable observables is useful for demonstrating steering~\cite{Quintino2014,Uola2014}, and every incompatibility problem can be mapped into a steering problem in a one-to-many manner~\cite{Uola2015,Uola2017,Kiukas2017}.

The extension of steering to multipartite systems has also been proposed. In the multipartite setting the concept of steering has some ambiguity in it. Whether one is interested in the typical spooky action at a distance~\cite{He2011,Cavalcanti2011,He2013} or in a more detailed semi-device independent entanglement verification scheme~\cite{Cavalcanti2015a,Riccardi2017}, one ends up with two different definitions. Here we are interested in the latter scenario as it relates more closely to our approach. 

\textls[-15]{To detect steerability of a given bipartite quantum state might turn into a cumbersome task. The~question of steerability (with given measurements on Alice's side and tomography on Bob's side) can be formulated as a semidefinite program (SDP)~\cite{pusey,skrypzikreview,Kogias2015} and as such one could imagine that the task is straightforward and easy to implement. Whereas SDP methods provide a powerful tool for steering detection, they are often restricted to systems with only a few measurements and small dimensions due to computational limitations. One should mention, though, that SDP methods can be used to set bounds for steering even in a scenario with a continuum of measurements~\cite{Fillettaz2018,Hirsch2016,Cavalcanti2016}. Another way of detecting steering is through criteria based on correlations~\cite{wiseman,ECavalcanti2015,Jevtic2014,chau2016,Bowles2016,Moroder2016}. Whereas these criteria are mostly analytical and straightforward to evaluate, they are also often either not optimal or limited to qubit systems.}

In Ref.~\cite{Costa2018}, steering criteria are developed from entropic uncertainty relations (EURs). The criteria are based on generalized entropies, hence, forming an extension of the entropic criteria in Refs.~\cite{walborn11, schneeloch13}. The work falls into the second category of the aforementioned classification of steering criteria, and, as~pointed out by the authors, the criteria of Ref.~\cite{Costa2018} manage to beat other correlation-based methods either in applicability or in detection power. In this work we extend the analysis of Ref.~\cite{Costa2018} to R\'enyi entropies and to tripartite steering scenarios. We discuss in detail the question of steerability with local and global measurements in the tripartite setting. Please note that recently similar efforts have been pursued in the context of R\'enyi entropies~\cite{Krivachy2018}.

This work is organized as follows. First, we introduce the concept of steering for bipartite and tripartite systems in Section~\ref{sec-steering}. Second, we present some useful entropies for the characterization of steering, followed by bounds of EURs in Section~\ref{sec-entropy}, where we also propose some bounds for Tsallis entropies, obtained from numerical investigations. We explain the criteria for the detection of steering from EURs in Section~\ref{sec-esc}. In Section~\ref{sec-otfried-huang} we provide a connection to existing entanglement criteria. In~Section~\ref{sec-appl} we investigate the optimal parameters from generalized entropies for the detection of steering, followed by the application of the criteria to some common examples. Finally, in Section~\ref{sec-multi} we extend the criteria to the tripartite case, and apply it to noisy GHZ and W states. We conclude the paper with some final remarks.
 

\section{Steering} 
\label{sec-steering}
In a bipartite steering scenario, Alice and Bob share a quantum state, Alice performs local actions (measurements) on her part of the state and Bob is left with non-normalised states (or a state assemblage) depending on Alice's choice of measurement and her reported outcomes. The task for Bob is to verify if his assemblages could be prepared using a separable state or not \cite{Moroder2016}. In a more formal manner, we can assume that Alice performs a measurement $A$ with outcome $i$ on her part of the system, while Bob performs a measurement $B$ with outcome $j$ on his part. From that, they can obtain the joint probability distribution of the outcomes. If for all possible measurements $A$ and $B$ one can express the joint probabilities in the form
\be
\label{lhs-model}
p(i,j|A,B) = \sum_\lambda p(\lambda)p(i|A,\lambda){ p_Q}(j|B,\lambda),
\ee
then the shared state is called unsteerable. Here, $p(i|A,\lambda)$ is a general
probability distribution, while ${ p_Q}(j|B,\lambda) = \textrm{Tr}_B[B(j) \sigma_\lambda]$
is a probability distribution originating from a quantum state $\sigma_\lambda$. 
Furthermore, $B(j)$ denotes a measurement operator, i.e., $B(j)\geq 0$ and 
$\sum_j B(j) = \mathbbm{1}$, and $\sum_\lambda p(\lambda) = 1$, where $\lambda$ 
is a label for the hidden quantum state $\sigma_\lambda$. A model as in
Equation~(\ref{lhs-model}) is called a local hidden state (LHS) model, and if it
exists, Bob can explain all the results through a set of local states $\{\sigma_\lambda\}$
which is only altered by the classical information about Alice's performed measurement and the recorded outcome. Otherwise, the state is called steerable. One should notice that for a state to be unsteerable, one has to prove the existence of an LHS model for all possible measurements on Alice's side and for a tomographically complete set on Bob's side, whereas for proving steerability it suffices to find a set of measurements for Alice and Bob for which the probabilities cannot be expressed as Equation~(\ref{lhs-model}).

For multipartite systems, LHS models can be extended in different ways. For simplicity, let us consider the case of tripartite systems. In addition to the notation used before, we assume that Charlie performs measurements $C$ with outcomes labelled with $k$. Then, one possibility is to ask if Alice can steer the state of Bob and Charlie. If for all possible measurements $A$, $B$ and $C$ the joint probability distribution can be expressed as
\be
\label{lhs-model-1-2}
p(i,j,k|A,B,C) = \sum_\lambda p(\lambda)p(i|A,\lambda){ p_Q}(j|B,\lambda){ p_Q}(k|C,\lambda),
\ee
the system is called unsteerable from Alice to Bob and Charlie. Here, ${ p_Q}(j|B,\lambda){ p_Q}(k|C,\lambda) = \textrm{Tr} [B(j)\otimes C(k)(\sigma_\lambda^B\otimes\sigma_\lambda^C)]$, where the hidden states of Bob and Charlie are factorizable. We require the factorizability in order to distinguish the tripartite scenario from a bipartite one (i.e., Bob and Charlie being a single system) where unsteerability is defined as
\be
\label{lhs-model-1-2-ent}
p(i,j,k|A,B,C) = \sum_\lambda p(\lambda)p(i|A,\lambda){ p_Q}(j,k|B,C,\lambda),
\ee
with ${ p_Q}(j,k|B,C,\lambda) = \textrm{Tr} [B(j)\otimes C(k)\sigma_\lambda^{BC}]$. Please note that the factorizability requirement includes all hidden state models using separable states through a redefinition of the hidden variable space. From a physical point of view, Equation~\eqref{lhs-model-1-2} corresponds to tests of full separability with untrusted Alice; whereas Equation~\eqref{lhs-model-1-2-ent} corresponds to tests of biseparability in the $A|BC$ cut with untrusted Alice.

Another possibility is to ask whether the joint probability distribution of measurements performed by Alice, Bob and Charlie, can be expressed as
\be
\label{lhs-model-2-1-local}
p(i,j,k|A,B,C) = \sum_\lambda p(\lambda)p(i|A,\lambda)p(j|B,\lambda){ p_Q}(k|C,\lambda),
\ee
which means that the system is unsteerable from Alice and Bob to Charlie with factorizable post-processing, meaning that we assume the post-processings on one party to be independent of that of the other party. This extra assumption is one possibility to distinguish, between bipartite and tripartite scenarios. Please note that one could also require non-signalling instead of factorizability of the post-processings. In a purely bipartite scenario an unsteerable joint probability distribution would be given by
\be
\label{lhs-model-2-1-global}
p(i,j,k|A,B,C) = \sum_\lambda p(\lambda)p(i,j|A,B,\lambda){ p_Q}(k|C,\lambda).
\ee
Similarly to the above scenario, Equation~\eqref{lhs-model-2-1-local} corresponds to tests of full separability with untrusted Alice and Bob; whereas Equation~\eqref{lhs-model-2-1-global} corresponds to tests of biseparability in the $AB|C$ cut with untrusted Alice and Bob.

One should notice that, in the steering scenario from Alice and Bob to Charlie, there is a difference whether Alice and Bob decide to perform global or local measurements. A simple example of this difference can be explored in the framework of super-activation of steering~\cite{super-activation}. Here, the authors show that while one copy of a quantum state is unsteerable, many copies of the same state become steerable, in the sense that steerability is ``activated''. Namely, consider a state $\varrho_{ABCC'} = \varrho_{AC}\otimes \varrho_{BC'}$, where $\varrho_{BC'}$ is a copy of $\varrho_{AC}$, and $\varrho_{AC}$ is unsteerable, but its steerability can be super-activated (where only two copies is already enough~\cite{super-activation}). For this state, local measurements give an unsteerable state assemblage, whereas, because of super-activation, it is steerable with global measurements.

\section{Entropies and Entropic Uncertainty Relations}
\label{sec-entropy}
\vspace{-6pt}

\subsection{Entropies}
\textls[-15]{Let us state some basic facts about entropies. For a general 
probability distribution \mbox{$\PP = (p_1,\dots,p_N)$}, the Shannon 
entropy is defined as~\cite{coverthomas} }
\be
S(\PP) = - \sum_i p_i\ln (p_i).
\ee

As a possible generalized entropy, we consider the so-called 
Tsallis entropy~\cite{Havrda1967,Tsallis1988} which depends 
on a parameter $0<q \neq 1$. It is given by
\be
\label{tsallis}
S_q (\PP) = - \sum_i p_i^q \ln_q (p_i),
\ee
where the $q$-logarithm is defined as $\ln_q(x) = ({x^{1-q}-1})/({1-q})$. 
Another generalization of Shannon entropy is known as R\'enyi entropy~\cite{Renyi1966}, which is defined depending on a parameter $0<r\neq 1$ as 
\be
{ \tilde{S}_r} (\PP) = \frac{1}{1 - r}\ln\left[\sum_i p_i^r\right].
\ee

The above entropies have the following properties~\cite{coverthomas,Havrda1967,Tsallis1988,Renyi1966}:

\begin{enumerate}[leftmargin=*,labelsep=4.9mm]
\item 	The entropies $S,S_q$ and ${ \tilde{S}_r}$ are positive and they are zero if and only if the probability distribution is concentrated at one value ($k$), i.e., $p_i = \delta_{ik}$.
\item	In the limit of $q \rightarrow 1$ and $r \rightarrow 1$, the Tsallis and 
R\'enyi entropies converge to the Shannon entropy, and both decrease monotonically in $q$ and $r$.
\item	The R\'enyi entropy is a monotonous function of the Tsallis entropy:
\be
{ \tilde{S}_r} (\PP) = \frac{\ln[1+(1-r)S_{ q=r} (\PP)]}{1-r}.
\ee
\item Shannon and Tsallis entropy are concave functions in $\PP$, i.e., they obey the relation
\be
f(\lambda \PP_1 + (1-\lambda)\PP_2) \geq \lambda f(\PP_1) + (1-\lambda)f(\PP_2),
\ee
where $f = S$ for Shannon entropy, and $f = S_q$ for Tsallis entropy. The R\'enyi entropy is concave if $r\in(0;1)$, and for other values of $r$ it is neither convex nor concave.
\item 	In the limit of $r\rightarrow \infty$, the R\'enyi entropy is known as min-entropy
\be
\lim_{r\rightarrow\infty} { \tilde{S}_r}(\PP) = -\ln\max_i (p_i).
\ee
\item For two independent distributions, $\PP$ and $\QQ$, Shannon and R\'enyi entropies are additive, i.e.,
\begin{align}
S(\PP, \QQ) = S(\PP)+S(\QQ), \\
{ \tilde{S}_r}(\PP, \QQ) = { \tilde{S}_r}(\PP)+{ \tilde{S}_r}(\QQ),
\end{align}
whereas Tsallis entropy is pseudo-additive, i.e.,
\be
S_q(\PP, \QQ) = S_q(\PP)+S_q(\QQ) + (1-q)S_q(\PP)S_q(\QQ).
\ee
\end{enumerate}

\subsection{Relative Entropies}

The relative entropy, also known as 
Kullback--Leibler divergence~\cite{coverthomas}, for two 
probability distributions $\PP$ and $\QQ$ is given by
\be\label{eq-kld}
D(\PP||\QQ) = \sum_i p_i \ln \Big(\frac{p_i}{q_i}\Big).
\ee 

For Tsallis and R\'enyi entropies the relative entropy is defined as~\cite{Tsallis1998,Furuichi2004,Renyi1966}
\be
D_q (\PP||\QQ) = 
-\sum_i p_i \ln_q \Big(\frac{q_i}{p_i}\Big), \qquad
{ \tilde{D}_r} (\PP ||\QQ) = \frac{1}{r - 1}\ln\left(\sum_i p_i^r q_i^{1-r}\right),
\ee
respectively. The R\'enyi relative entropy is also known as R\'enyi divergence.

Here, we discuss two properties which are essential in this work: first, the relative entropy is additive for independent distributions, that is if $\PP_1,\PP_2$ 
are two probability distributions with the joint distribution 
$\PP(x,y) = \PP_1(x)\PP_2(y)$, and the same for $\QQ_1,\QQ_2$, then one has
\be
\label{eq-add}
D(\PP||\QQ) = D(\PP_1||\QQ_1) + D(\PP_2||\QQ_2),
\ee
and the same holds for the generalized R\'enyi relative entropy,
\be
{ \tilde{D}_r} (\PP||\QQ) = { \tilde{D}_r} (\PP_1||\QQ_1) + { \tilde{D}_r} (\PP_2||\QQ_2).
\ee

However, for the generalized Tsallis relative entropy, we have~\cite{Furuichi2004}
\be\label{eq-add-tsallis}
D_q (\PP|| \QQ) = D_q(\PP_1||\QQ_1) + D_q(\PP_2||\QQ_2) + (q-1)D_q(\PP_1||\QQ_1)D_q(\PP_2||\QQ_2),\nonumber
\ee
where the additional term is due to the pseudo-additivity of the generalized entropy.

Second, the relative entropy is jointly convex. This means that for two
pairs of distributions $\PP_1, \QQ_1$ and $\PP_2, \QQ_2$ one has
\be
D[\lambda \PP_1 +  (1-\lambda)\PP_2||\lambda \QQ_1 + (1-\lambda) \QQ_2] \leq \lambda D(\PP_1||\QQ_1) + (1-\lambda)D(\PP_2||\QQ_2).
\label{eq-jointconv}
\ee

The generalized Tsallis relative entropy is also jointly convex for all values of $q$, while the generalized R\'enyi relative entropy is jointly convex only for $r \in (0;1)$ (see Theorem 11 in Ref.~\cite{Erven2014}).

\subsection{Entropic Uncertainty Relations}

Entropies are useful for the investigation of uncertainty relations~\cite{Deutsch1983}. 
Entropic uncertainty relations (or EURs for short) can be easily explained with an example. 
Consider the Pauli measurements $\sigma_x$ and $\sigma_z$ on a single 
qubit. For any quantum state these measurements give rise to a two-valued probability
distribution and to the corresponding entropy $S(\sigma_m)$ for $m=x,z$.
The fact that $\sigma_x$ and $\sigma_z$ do not share a common eigenstate can be
expressed as~\cite{Maassen1988}
\begin{equation}\label{example-x-z}
S(\sigma_x) +  S(\sigma_z) \geq \ln(2), 
\end{equation}
where the lower bound does not depend on the state. These type of relations can be extended to more measurements and other entropies, and the search for the optimal bounds is an active field of research.

In a general way, if one performs $m$ measurements, the bounds of an EUR can be estimated in the following way
\be
\sum_m S(X_m) \geq \min_{\varrho} \sum_m S(X_m)_{\varrho} = \BB,
\ee
where the minimization, due to the concavity of the entropy, involves all pure (single system) states. Various analytical entropic uncertainty bounds are known for Shannon, Tsallis and R\'enyi entropies, and we introduce some of them in this section, together with new bounds for Tsallis entropy obtained from numerical investigations. These bounds will be useful in later sections, where we develop steering criteria based on the relative entropy between two probability distributions.

For the estimation of the bounds for EURs we consider mutually unbiased bases (MUBs)~\cite{Durt2010}. Two orthonormal bases are mutually unbiased if the absolute value of the overlap between any vector from one basis with any vector from the other basis is equal to $1/d$. For a given dimension $d$, it is simple to construct a pair of MUBs through, for example, the discrete Fourier transform. If $d$ is a prime or power of a prime, the existence of $d+1$ MUBs (i.e., a complete set of MUBs) is known. However, the number of MUBs existing in other than prime and power of prime dimensions is a long standing open problem~\cite{Bengtsson2007}.

For the Shannon entropy and a complete set of MUBs (provided that they exist), the bounds for EURs for dimension $d$ were analytically derived in~Ref.~\cite{Ruiz1995} and are given by~
\be
\label{boundq1}
\BB = 
\begin{cases}
(d+1)\ln \left(\frac{d+1}{2}\right), & \quad d\, \text{odd} \\ \\
\frac{d}{2}\ln \left(\frac{d}{2}\right) + \left(\frac{d}{2}+1\right)\ln \left(\frac{d}{2}+1\right), & \quad d\, \text{even}. \\
\end{cases}
\ee

Later, a bound was proved in Ref.~\cite{Wu2009} for $m$ MUBs (which coincides with the above bound for a complete set),
\be\label{bound-wu}
\BB = m \ln(K) + (K+1)\left(m-K \frac{d+m-1}{d}\right)\ln\left(1+\frac{1}{K}\right),
\ee
where $K = \left\lfloor \frac{md}{d+m-1}\right\rfloor$ and $\lfloor\cdot \rfloor$ is the floor function. Please note that steering with MUBs can be also attacked using techniques from the field of joint measurability~\cite{Erkka2015,Kimmo2016,Designolle2018}.

We are also interested in EURs not only for single systems, but for composite ones as well. 
For~bipartite systems we have the following bound
\be
\sum_m S(X^A_m,X^B_m) \geq \min_{\varrho_{AB}} \sum_m S(X^A_m,X^B_m)_{\varrho_{AB}} = \CC,
\ee
where the minimization involves all pure single system states. Here, $S(X^A_m,X^B_m)$ is the Shannon entropy of the probability distribution $p_{ij}^{(m)} = \bra{i_m}\bra{j_m}\varrho_{AB}\ket{i_m}\ket{j_m}$, with $d^2$ outcomes, where the MUBs $\{\ket{i_m}\}_m,\{\ket{j_m}\}_m$ work as the eigenvectors of the measurement $X^{A(B)}_m$. Please note that we use the symbol $\BB$ for the bounds on single systems, while $\CC$ is used for composite ones. 

One should note that there might be a difference between the bounds obtained from separable and entangled states. A simple example is given by a two-qubit system and Pauli measurements. Optimizing over all two-qubit states and considering two Pauli measurements, we have
\be\label{bound-s-sep}
S(\sigma_x,\sigma_x) + S(\sigma_y,\sigma_y) \geq 2\ln(2).
\ee

This bound is already reached with separable states~\cite{Rene2018}. Meanwhile, if one considers separable states and three Pauli measurements (which represent a complete set of MUBs for two-dimensional systems), the bound is
\be\label{bound-s-3-sep}
\sum_m S(\sigma_m,\sigma_m)_{\varrho_{sep}} \geq 4\ln(2),
\ee 
where $m= \{x,y,z\}$. However, if one considers the maximally entangled state $\varrho_{ent} = \ket{\psi^-}\bra{\psi^-}$, the~following value is reached
\be\label{bound-s-3-ent}
\sum_m S(\sigma_m,\sigma_m)_{\varrho_{ent}} = 3\ln(2),
\ee
for the same measurements. Please note that, for separable states, the bound in Equation~\eqref{bound-s-sep} follows from additivity of Shannon entropy and Equation~\eqref{example-x-z}, whereas the bound in Equation~\eqref{bound-s-3-sep} follows from additivity and the bound in Ref.~\cite{Ruiz1995}. Moreover, the additivity of EUR for Shannon entropy is discussed in Ref.~\cite{Rene2018}.

An analytical bound for separable states ($\varrho = \sum_j p_j \varrho_j^A\otimes \varrho_j^B$, with Hilbert space dimensions $d_A$ and $d_B$) and $m$ MUBs performed in each system is given by~\cite{Wu2009}

\begin{eqnarray}\label{bound-sep-general}
\CC &=& m\ln (K_A) + m\ln (K_B) + (K_A + 1)\left(m - K_A \frac{d_A + m -1}{d_A}\right)\ln\left(1 + \frac{1}{K_A}\right) \nonumber \\
&& + (K_B + 1)\left(m - K_B \frac{d_B + m -1}{d_B}\right)\ln\left(1 + \frac{1}{K_B}\right), \nonumber \\
\end{eqnarray}
with $K_{A(B)}$ defined as above. Please note that this bound is the sum of the bounds~\eqref{bound-wu} for both subsystems. Here, this bound also holds because of concavity and additivity of Shannon entropy.

Now, let us present the bounds for generalized entropies. For the Tsallis entropy and $m$ MUBs it has been shown in Ref.~\cite{Rastegin2013} that, for $q\in (0;2]$, the bound is given by 
\be
\label{boundq2}
\BB^{(q)} = m\ln_q \left(\frac{m d}{d + m -1}\right).
\ee

For $q\rightarrow 1$, this bound is not optimal 
for even dimensions, so in this case it is more appropriate to consider 
the bounds given in Equation~\eqref{boundq1}.

In Ref.~\cite{Guhne2004}, a bound for the Tsallis entropy and two-qubit systems was analytically derived, and for every $q\in [2n-1,2n], n\in\mathbb{N}$, the bound is 
\be\label{bound-t-qubit}
\BB^{(q)} = \ln_q(2), 
\ee
for two-measurement settings composed by Pauli operators, which are MUBs in dimension 2. Numerically, these bounds seem also to hold for other values of $q$, except $q\in (2;3)$. 
For three measurement settings, one can obtain numerically the following bound
\be
\BB^{(q)} = 2\ln_q(2),
\ee
which is also not optimal for $q\in (2;3)$ (see Ref.~\cite{Guhne2004}). Extending these bounds for arbitrary (finite) dimensions and $m$ mutually unbiased measurements, numerical investigations suggest that, for $q\geq 2$, 
\be\label{bound-t-general}
\BB^{(q)} = (m-1)\ln_q(d).
\ee

To be more precise, the above function seems to match the numerically calculated optimal values for small values of $q$, $d$ and $m$.

\begin{figure}[H]
\centering
\includegraphics[width=7.5 cm]{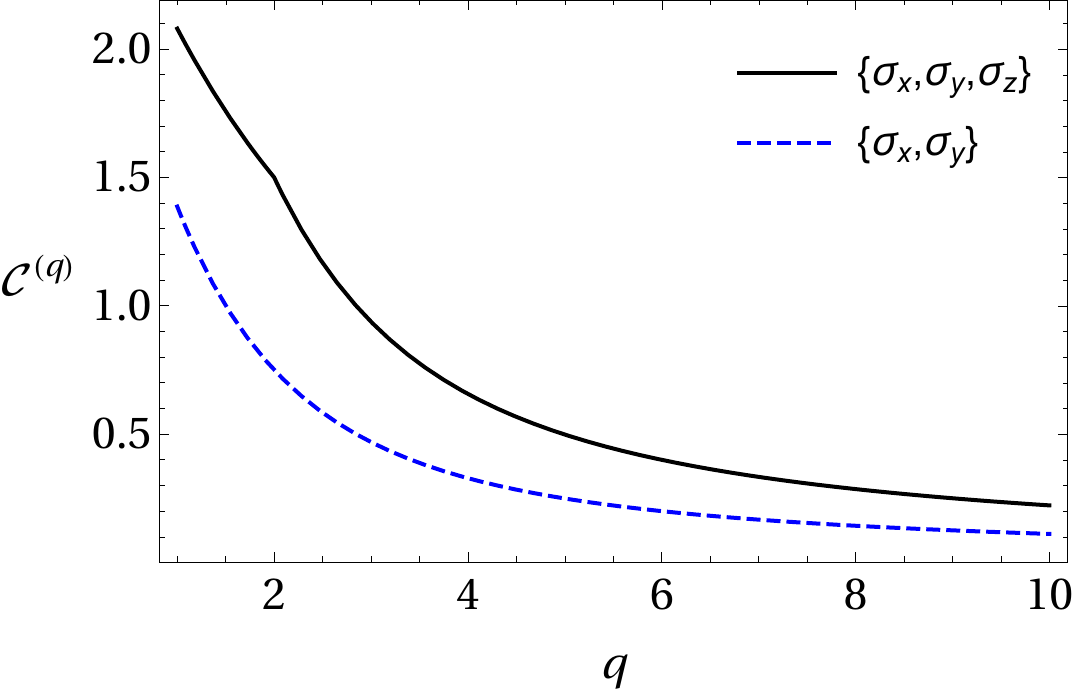}
\caption{Numerical lower bounds for composite systems in Equations \eqref{bound-q-2} and \eqref{bound-q-3} in terms of the parameter~$q$.}
\label{fig-bound-sep-ent}
\end{figure}

Now, if one considers two-qubit systems, we introduce here the bounds for Tsallis entropy (with $q>1$), obtained from numerical investigation by minimizing over all pure states. They are given by
\be\label{bound-q-2}
\CC^{(q)} = \ln_q(4)
\ee
for two Pauli measurements, where this bound is already reached by separable states, and
\be\label{bound-q-3}
\CC^{(q)} = 
\begin{cases} 3\ln_q(2), &  1\leq q \leq 2 \\
2\ln_q(4) & q \geq 2
\end{cases}
\ee
for three Pauli measurements. Here, in the range of $1\leq q \leq 2$ the bound gets lower due to entanglement. In the range $q\geq 2$ separable states give the best bounds for this setting of measurements. In Figure~\ref{fig-bound-sep-ent} we show these results. All these bounds were obtained numerically and as their analytical proof remains an open question, we use these conjectured bounds in our calculations.

Regarding the bounds for R\'enyi entropy, we have the following scenarios~\cite{Rastegin2013}: in the range $r\in (0;2)$ the bounds are independent of $r$ and they equal the bounds for Shannon entropy; and for $r\in [2;\infty)$ the state-independent bounds are
\be
{ \tilde{\BB}_r}^{(r)} = \frac{m r}{2(r -1)}\ln\left(\frac{m d}{d+m-1}\right).
\ee

\section{Entropic Steering Criteria}
\label{sec-esc}

In this section we present the detailed derivation of the generalized entropic steering criteria proposed in Ref.~\cite{Costa2018}. Here we show the results for Shannon and Tsallis entropies, as has been made in Ref.~\cite{Costa2018}, and also extend the criteria for R\'enyi entropy. Please note that the proof is not at all restricted to these functions as it can be applied to all functions which satisfy the following properties: (i)~(pseudo-)additivity for independent distributions; (ii) state independent EUR; and~(iii)~joint convexity of the relative entropy. In the following we present our proof for specific entropies, and the method becomes clear from its application to each of them.

\subsection{Entropic Steering Criteria for Shannon Entropy}

\textls[-20]{The starting point of our method is to consider the relative entropy~\eqref{eq-kld} between \mbox{two distributions, i.e.,}}

\be
F(A,B) = - D(A\otimes B||A \otimes \mathbb{I}).
\ee

Here $A\otimes B$ denotes the joint probability distribution 
$p(i,j|A,B)$, which we further denote by $p_{ij}$, $A$ is the marginal distribution
$p(i|A)$, which we denote by $p_{i}$, and $\mathbb{I}$ is a uniform distribution with $q_j = 1/N$ for all outcomes $j \in \{1,\cdots ,N\}$. As the relative entropy is 
jointly convex, $F(A,B)$ is concave in the probability distribution 
$A\otimes B$. Then, we get
\begin{equation}
\label{eq-res1}
F(A,B) = 
- \sum_{ij} p_{ij} \ln\Big(\frac{p_{ij}}{p_i/N}\Big) = S(A,B) - S(A) - \ln(N) = S(B|A) - \ln(N),
\end{equation}
where $S(B|A)$ is the Shannon conditional entropy. 
On the other hand, considering a product distribution $p(i|A,\lambda){ p_Q}(j|B,\lambda)$ with a fixed $\lambda$ and using the property from Equation~(\ref{eq-add}), 
we have
\begingroup\makeatletter\def\f@size{9}\check@mathfonts
\def\maketag@@@#1{\hbox{\m@th\fontsize{10}{10}\selectfont\normalfont#1}}%
\be
F^{(\lambda)}(A,B) = - D[p(i|A,\lambda)||p(i|A,\lambda)] 
- D[{ p_Q}(j|B,\lambda)||\mathbb{I}] = - D[{ p_Q}(j|B,\lambda)||\mathbb{I}] = S^{(\lambda)}(B) - \ln(N).
\label{eq-res2}
\ee
\endgroup

The term $S^{(\lambda)}(B)$ on the right-hand side of this equation depends on probability 
distributions taken from the quantum state $\sigma_\lambda$. 
For a given set of measurements $\{B_m\}$, 
such distributions typically obey an EUR
\begin{equation}
\sum_m S^{(\lambda)}(B_m) \geq \BB_B,
\label{eq-eur}
\end{equation}
\textls[-20]{where $\BB_B$ is some entropic uncertainty bound for the observables $B_m$. 
Finally, since $S$ is concave, the same bound holds for convex 
combinations of product distributions $p(i|A,\lambda){ p_Q}(j|B,\lambda)$ from Equation~(\ref{lhs-model}). Connecting this to Equations~\eqref{eq-res1} and~\eqref{eq-res2} we have, 
for a set of measurements \mbox{$\{A_m\otimes B_m\}_m$},}
\be
\label{eq-res3}
\sum_m S(B_m|A_m) \geq \BB_B,
\ee
which means that any nonsteerable quantum system obeys this relation.
In this way EURs can be used to derive
steering criteria. The intuition behind these criteria is based on the 
interpretation of Shannon conditional entropy. In Equation~(\ref{eq-res3}), one can see 
that the knowledge Alice has about Bob's outcomes
is bounded. If this inequality is violated, then the system is steerable, meaning 
that Alice can do better predictions than those allowed by an EUR.

This criterion is more general than the one in Ref.~\cite{schneeloch13}, since our proof can easily also be extended to other generalized entropies, as we show in the following.

\subsection{Entropic Steering Criteria for Generalized Entropies}
\vspace{-6pt}

\subsubsection{Tsallis Entropy}

\textls[-15]{Now we can apply the machinery derived above and consider the 
quantity \mbox{$F_q(A,B) = -D_q(A\otimes B||A\otimes \mathbbm{I})$}.
Using the definition of the generalized relative entropy, we have}
\be\label{tsallis-1}
F_q (A,B) = \sum_{i,j}p_{ij}\ln_q \left(\frac{p_i/N}{p_{ij}}\right) = \frac{x}{1-q} + (1+x)\Big\{S_q(B|A)+(1-q)C(A,B)\Big\},
\ee
where $S_q(B|A) = S_q(A,B) - S_q(A)$ is the conditional Tsallis 
entropy~\cite{Furuichi2006}, $x = N^{q-1}-1$, and
\begin{equation}\label{tsallis-2}
C(A,B) = \sum_ip_i^q [\ln_q(p_i)]^2 
- \sum_{i,j}p_{ij}^q \ln_q (p_i)\ln_q (p_{ij}),
\end{equation}
is the correction term.

Now, considering the property from Equation~(\ref{eq-add-tsallis}) and a product distribution $p(i|A,\lambda){ p_Q}(j|B,\lambda)$ with a fixed $\lambda$ one gets
\begin{eqnarray}
F_q^{(\lambda)}(A,B) = \frac{x}{1-q} + (1+x)S_q^{(\lambda)}(B).
\label{eq-res2-t}
\end{eqnarray}

It follows by direct calculation that if the measurements $\{B_m\}_m$ 
obey an EUR
\be
\sum_m S_q(B_m)\geq \BB^{(q)}_B
\ee
then one has the steering criterion
\be
\label{tsc}
\sum_m\Big[S_q(B_m|A_m)+(1-q)C(A_m,B_m)\Big]
\geq \BB_B^{(q)}.
\ee

From Equation~(\ref{tsc}) it is easy to see that if we consider $q\rightarrow 1$, we arrive 
at Equation~(\ref{eq-res3}). Note~that one can rewrite Equation~\eqref{tsc} 
in terms of probabilities as
\begin{equation}
\label{tsc-prob}
\frac{1}{q-1}\left[\sum_{k}\left(1 - \sum_{ij}\frac{(p_{ij}^{(m)})^q}{(p_{i}^{(m)})^{q-1}}\right)\right] \geq \BB_B^{(q)}.
\end{equation}

Here, $p_{ij}^{(m)}$ is the probability of Alice and Bob for outcome $(i,j)$ 
when measuring $A_m\otimes B_m$, and $p_i^{(m)}$ are the marginal outcome probabilities of Alice's measurement $A_m$. This form of the criterion
is straightforward to evaluate.

\subsubsection{R\'enyi Entropy}

If one considers the quantity ${ \tilde{F}_r}(A,B) = -{ \tilde{D}_r} (A\otimes B||A\otimes \mathbbm{I})$ with the measurements $B_m$ obeying an EUR
\be
\sum_m { \tilde{S}_r} (B_m) \geq { \tilde{\BB}_B^{(r)}},
\ee
we have the following steering criterion for R\'enyi entropy
\be\label{tsr}
\frac{1}{1-r}\sum_m \ln \left[\sum_{i,j}(p_{ij}^{(m)})^r\, (p_i^{(m)})^{1-r}\right] \geq { \tilde{\BB}_B^{(r)}}.
\ee

Please note that for the range $r \in (0;1)$ the bound is independent of $r$, and it is the same as the bound for Shannon entropy~\cite{Rastegin2013}. Unlike the other entropies, we cannot write the result in terms of R\'enyi conditional entropies, given its definition is not clear in the literature (see discussion in Ref.~\cite{Fehr2014}).

\section{Connection to Existing Entanglement Criteria}
\label{sec-otfried-huang}

At this point, it is interesting to connect our approach with
the entanglement criteria derived from EURs~\cite{Guhne2004}. In Ref.~\cite{Guhne2004}, it has been shown that for separable states the following inequality
\be
\label{eq-criteria-sep}
S_q(A_1\otimes B_1) + S_q(A_2\otimes B_2) \geq \BB_B^{(q)}
\ee
holds. Here, $A_1$ and $A_2$ ($B_1$ and $B_2$) are observables on Alice's (Bob's) 
laboratory, and Bob's observables obey an EUR $S_q(B_1) + S_q(B_2) \geq \BB_B^{(q)}$. 
Differently from our approach, $S_q(A_m\otimes B_m)$ is the entropy of the probability
distribution of the outcomes of the {\it global} observable 
$A_m\otimes B_m$. Please note that for a degenerate
$A_m\otimes B_m$ the probability distribution differs from the 
local ones. For instance, measuring $\sigma_z \otimes \sigma_z$
gives four possible local probabilities $p_{++}, p_{+-}, p_{-+}, p_{--},$
but for the evaluation of $S(A_m\otimes B_m)$ one combines them
as $q_+ = p_{++}+p_{--}$ and  $q_- = p_{+-}+p_{-+}$, as these
correspond to the global~outcomes. 

There are some interesting connections between our derivation of steering inequalities and this entanglement criterion. 
First, the proof in Ref.~\cite{Guhne2004} is based on EURs for Bob's observables (the same as our criteria), and this is the only quantum restriction in the criterion, so Equation~(\ref{eq-criteria-sep}) is a steering inequality, meaning that all probability distributions of the form in Equation~(\ref{lhs-model}) fulfil it.
Second, in~Ref.~\cite{Guhne2004} it was observed that the criterion is
strongest for values $2 \leq q \leq 3$, which seems to be the case also for our criteria (shown later). Third, for special scenarios (e.g., Bell-diagonal two-qubit states and Pauli measurements), Equation~(\ref{eq-criteria-sep}) and Equations~(\ref{eq-res3}) and (\ref{tsc}) give the same results. However, it~does not hold for more general scenarios.

The approach of Ref.~\cite{Guhne2004} has been slightly improved in Ref.~\cite{Huang2010}, where the main idea is to recombine the probability distribution in a different way (see below). Also, the criteria in Ref.~\cite{Huang2010} are more general in the sense that can be applied to any symmetric and concave function. 

Similar to the case of Ref.~\cite{Guhne2004}, the criteria from Ref.~\cite{Huang2010} can be also applied to steering. To see this, let us first explain the main ideas in~\cite{Huang2010}. In this work, they consider concave and symmetrical (i.e., invariant under the permutation of variables) functions $f:\mathbb{R}^n \longrightarrow \mathbb{R}$. 
For simplicity, define
\be
f(\rho_0,e) = f(\bra{e_1}\rho_0\ket{e_1},\dots,\bra{e_n}\rho_0\ket{e_n}),
\ee
where $e=\{\ket{e_i}|i=1,\dots,n\}$ is an orthonormal basis of the $n$-dimensional Hilbert space in which the state $\rho_0$ acts. Then, one can construct a probability matrix $P=(p_{ij})$, where the elements are defined as $p_{ij} = \bra{e_i^A}\bra{e_j^B}\rho\ket{e_j^B}\ket{e_i^A}$, where $e_{A} = \{\ket{e_i^A}|i=1,\cdots,n_A\}$ and $e_{B} = \{\ket{e_j^B}|j=1,\cdots,n_B\}$ are orthonormal bases of the $n_{A(B)}$-dimensional Hilbert space $\mathcal{H}_{A(B)}$. 

Then, define a permutation matrix $Q = (q_{ij})$, where $\{q_{i1},\dots,q_{in_B}\}$ is a permutation of an $n_B$-element set $S=\{s_1,\dots,s_{n_B}\}$ for $i=1,\dots,n_A$. For example, if we consider the case of $n_A = n_B = 3$, three examples of possible constructions of $Q$ are
\be\label{eq-examples-q}
\left(\begin{array}{ccc}
s_1 & s_1 & s_1 \\
s_2 & s_2 & s_2 \\
s_3 & s_3 & s_3
\end{array}
\right), \qquad
\left(\begin{array}{ccc}
s_1 & s_1 & s_3 \\
s_2 & s_3 & s_1 \\
s_3 & s_2 & s_2
\end{array}
\right), \qquad
\left(\begin{array}{ccc}
s_3 & s_1 & s_2 \\
s_1 & s_3 & s_3 \\
s_2 & s_2 & s_1
\end{array}
\right).
\ee

Now, define
\begin{equation}\label{huang}
f(\varrho_{AB},e_A,e_B,Q) = f\left(\sum_{ij}\delta(q_{ij},s_1)p_{ij},\sum_{ij}\delta(q_{ij},s_2)p_{ij},\cdots,\sum_{ij}\delta(q_{ij},s_{n_A})p_{ij} \right),
\end{equation} 
where $\delta(a,b)$ is the Kronecker function. Here, the argument of this function is the combination of the probabilities given a permutation matrix $Q$. If we take the third example in Equation~\eqref{eq-examples-q}, we have
\be
f(\varrho_{AB},e_A,e_B,Q) = f(p_{21}+p_{12}+p_{33},p_{31}+p_{32}+p_{13},p_{11}+p_{22}+p_{23}).
\ee

\textls[-15]{Here one can see that the combination of the probabilities will depend on the permutation matrix $Q$.} 

\textls[-20]{Given the above definitions, the authors prove the following bound for product states (\mbox{$\varrho_{AB} = \varrho_A\otimes\varrho_B$}),}
\be\label{eq-lemma-huang}
f(\varrho_{AB},e_A,e_B,Q) \geq f(\varrho_B,e_B),
\ee
\textls[-20]{holding for any permutation matrix $Q$ and any concave symmetrical function $f$. The bound is an entanglement criterion for pure states. Here, one can notice that the right-hand side of Equation~\eqref{eq-lemma-huang} is independent of the space $\mathcal{H}_A$, giving some hint that the criterion actually detects steerability of the state.} 

Using the notation $e_k^A = \{\ket{e_{i_k}^A}|i=1,\dots,n_A\}$ and $e_k^B = \{\ket{e_{j_k}^B}|j=1,\dots,n_B\}$ for different bases of $\mathcal{H}_{A(B)}$, the authors prove that for any separable state $\varrho_{AB}$
\be\label{eq-theorem-huang}
\sum_k f_k(\varrho_{AB},e_k^A,e_k^B,Q_k) \geq \min_{|\psi\rangle\in\mathcal{H}_B}\sum_k f_k(|\psi\rangle\langle\psi|,e_k^B)
\ee
holds for arbitrary symmetrical concave functions $f_k$, permutation matrices $Q_k$ and bases $e_k^{A(B)}$. Equation~\eqref{eq-theorem-huang} is a general entanglement criterion based on symmetrical concave functions $f_k$. In~order to find the optimal criteria, an optimization over all possible permutation matrices should be performed. A specific criterion is given for the case where $f_k$ is replaced by the Shannon entropy, and the bound in Equation~\eqref{eq-theorem-huang} is related to EURs. 

Now we show that the above entanglement criterion is actually a steering criterion, given that Equation~\eqref{eq-theorem-huang} can be obtained if one considers an LHS model.
Note first that one can include general measurements into the above considerations by defining
\begin{equation}
f(\varrho_0,M) := f(\textrm{Tr}[\varrho_0 M_i]_i), \quad\textrm{for}\quad i=1,\cdots ,n,
\end{equation}
where the operators $\{M_i\}_i$ form a positive operator valued measure (POVM) and $\varrho_0$ is a quantum state.
Then, taking an unsteerable state $\rho_{AB}$ and labelling by $N_i$ the POVM elements of Alice's measurements, one has for a fixed hidden variable $\lambda$
\begin{equation}\label{eq-lhs-huang}
p_{ij}(\lambda) = p(i|N,\lambda)\textrm{Tr}[M_j\rho^B_\lambda].
\end{equation}
Hence, 
\begin{eqnarray}
f\left(\sum_{ij}\delta(q_{ij},s_1)p_{ij}(\lambda),\cdots \right) &=& f\left(\sum_i p(i|N,\lambda)\sum_j \delta(q_{ij},s_1)\textrm{Tr}[M_j\rho^B_\lambda],\cdots\right) \nonumber \\
&\geq&  \sum_i p(i|N,\lambda)f\left(\sum_j \delta(q_{ij},s_1)\textrm{Tr}[M_j\rho^B_\lambda],\cdots\right)\nonumber\\
&=& \sum_i p(i|N,\lambda)f\left(\textrm{Tr}[M_1\rho^B_\lambda],\cdots\right)\nonumber\\
&=& f(\varrho^B_\lambda ,M).
\end{eqnarray}

On the second line we use concavity of the function $f$, and in the third line we use symmetry. Taking the sum over all hidden variables $\lambda$ gives
\be
f(\varrho,M,N,Q) := f\left(\sum_{ij}\sum_\lambda \delta(q_{ij},s_1)p(\lambda)p_{ij}(\lambda),\cdots\right) \geq \sum_\lambda p(\lambda)f(\varrho^B_\lambda,M) \geq \min_{\varrho \in \mathcal{H}_B} f(\varrho^B,M).
\ee

Considering more measurements one has
\be\label{eq-huang-steering}
\sum_k f_k(\varrho,M_k,N_k,Q_k) = \sum_\lambda p(\lambda)\sum_k f(\varrho^B_\lambda,M_k) \geq \min_{|\psi\rangle\in \mathcal{H}_B}\sum_k f_k(|\psi\rangle\langle\psi|,M_k),
\ee
which is exactly the same criteria of Equation~\eqref{eq-theorem-huang}. This means that the entanglement criteria proposed in Ref.~\cite{Huang2010} are actually steering criteria.


\section{Applications}
\label{sec-appl}
\vspace{-6pt}

\subsection{Optimal Values of $q$ and $r$ for Steering Detection}

\textls[-20]{In this section we investigate the dependence of our steering criteria on the parameters $q$ and $r$ appearing in Tsallis and R\'enyi entropies. Also a comparison between the criteria obtained from Tsallis and R\'enyi entropy (with Shannon entropy as a special case) is presented for specific examples. We~base our calculations on numerics for the cases where the optimal (analytical) uncertainty bounds are not known.}

Let us first consider the case of qubit systems. For this analysis, consider three noisy two-qubit entangled states, $\varrho^{(2)}_{e_x} (w) = w\rho^{(2)}_x + (1-w)\mathbbm{1}/4$ with $x=1,2,3$ where $\rho^{(2)}_1 = \ket{\psi^-}\bra{\psi^-}$, $\rho^{(2)}_2$ and $\rho^{(2)}_3$ are two example states,  given by
\begin{eqnarray}
\rho_2^{(2)} &=& \frac{1}{4}\left(
\begin{array}{cccc}
0.14 & 0.09 - 0.18i & -0.12+0.17i & -0.06 \\
0.09 + 0.18i & 1.58 & -1.72 & -0.12+0.17i \\
-0.12 - 0.17i & -1.72 & 1.98 & 0.09-0.18i \\
-0.06 & -0.12-0.17i & 0.09+0.18i & 0.3
\end{array}
\right), \nonumber \\
\rho_3^{(2)} &=& \frac{1}{4}\left(
\begin{array}{cccc}
0.06 & -0.13 & 0.16+0.02i & -0.02 \\
-0.13 & 1.74 & -1.82 & 0.16+0.02i \\
0.16 - 0.02i & -1.82 & 1.96 & -0.13 \\
-0.02 & 0.16-0.02i & -0.13 & 0.24
\end{array}
\right), \nonumber
\end{eqnarray}
which give a fair violation of the criteria. Please note that this behaviour is typical not only for these states. For all the states that we tried a similar plot was obtained. 

\begin{figure}[H]
\centering
\includegraphics[scale=0.5]{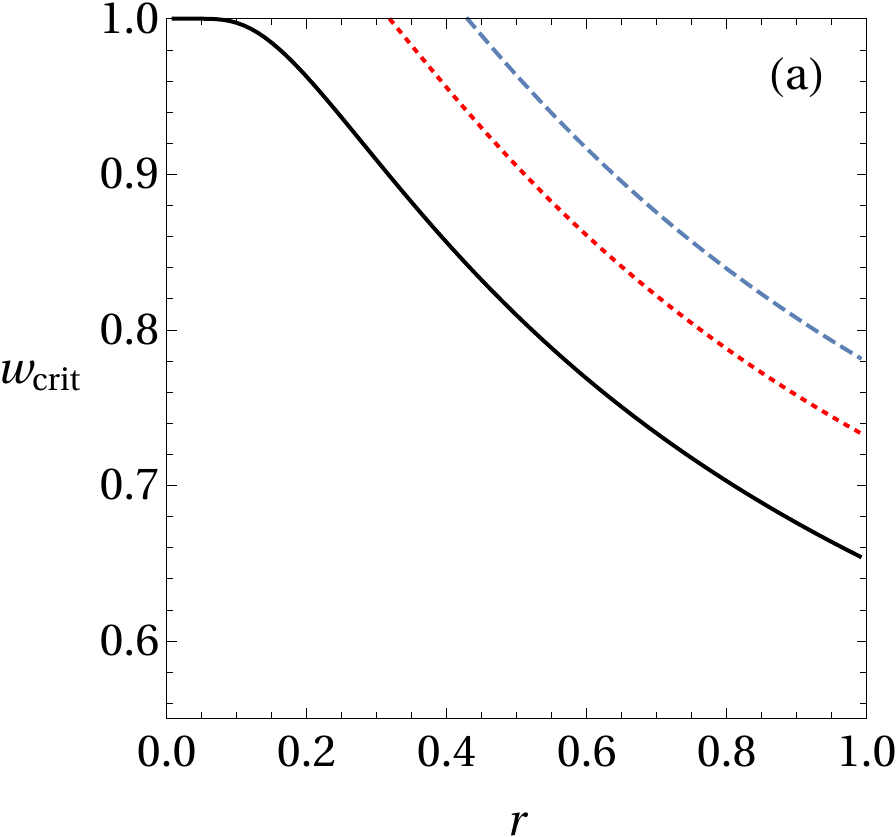}\includegraphics[scale=0.5]{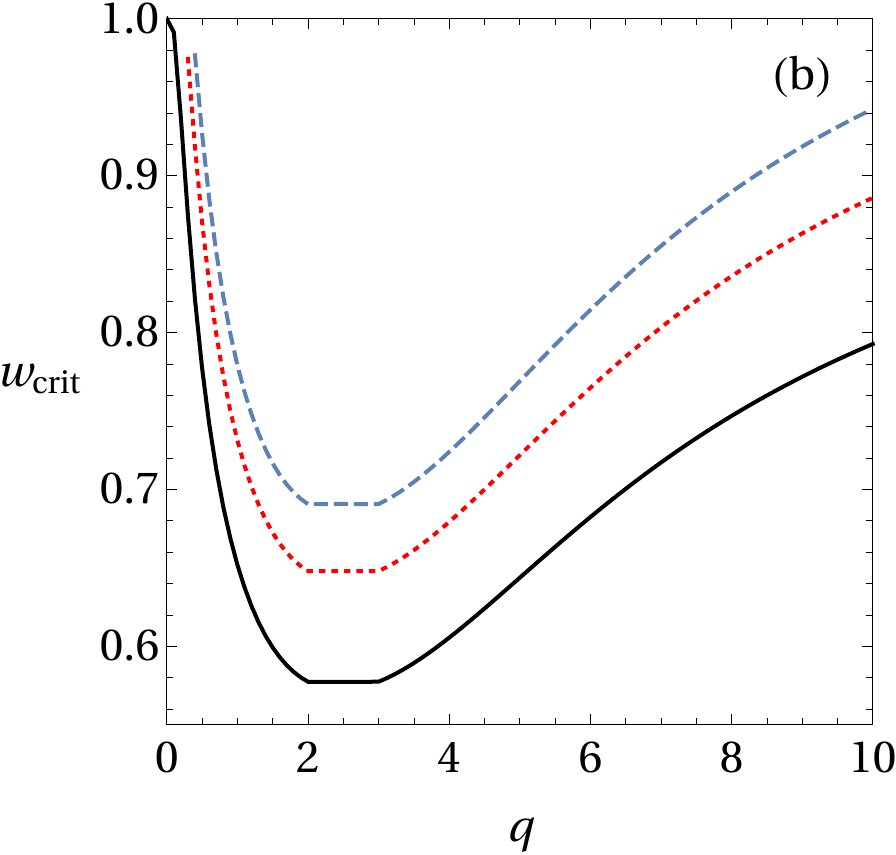}
\caption{The critical value $w$ for noisy two-qubit entangled states $\varrho^{(2)}_{e_x} (w)$ for the detection of steering. Solid black line corresponds to Werner states ({ $\varrho_{e_1}^{(2)}(w)$}), and the dashed blue and dotted red lines correspond to 
{ $\varrho_{e_2}^{(2)}(w)$ and $\varrho_{e_3}^{(2)}(w)$, respectively}, with (\textbf{a}) the criteria based on R\'enyi entropy [Equation~\eqref{tsc}] and (\textbf{b}) on Tsallis entropy [Equation~\eqref{tsr}].}
\label{fig-opt-q-1}
\end{figure}

Here we will focus on the Pauli measurements $\{\sigma_x,\sigma_y,\sigma_z\}$. In Figure~\ref{fig-opt-q-1} we show the critical value of white noise $w$ for the violation of the generalized entropic criteria from Equations~\eqref{tsc} and~\eqref{tsr}.

\begin{figure}[H]
\centering
\includegraphics[scale=0.6]{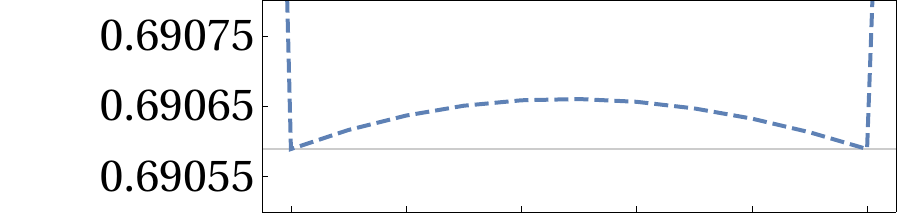}\\ \vspace{0.1cm}
\includegraphics[scale=0.6]{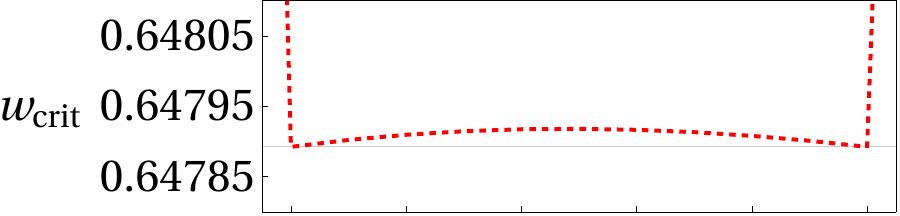}\\ \vspace{0.1cm}
\includegraphics[scale=0.6]{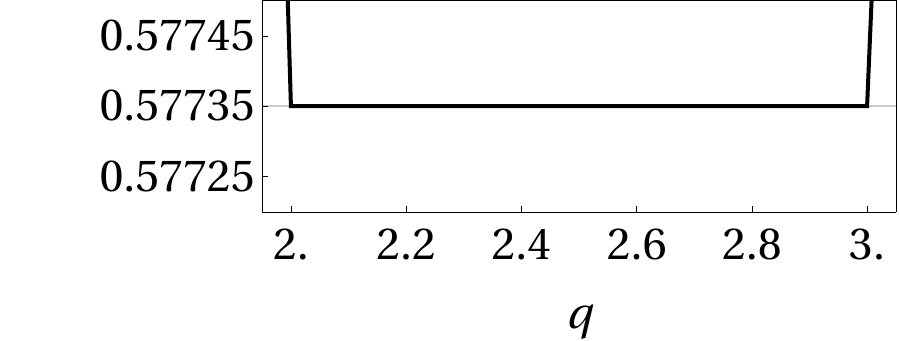}
\caption{Zoom-in of Figure~\ref{fig-opt-q-1}, for the interval $q\in [2;3]$. Solid black line corresponds to Werner states, and the dashed blue and dotted red lines correspond to two random entangled states.}
\label{fig-opt-q-2}
\end{figure}

From these simple examples, one is able to extract some hint about the optimal values of $q$ and $r$ that best identify steerability of the state. If one considers the criteria based on R\'enyi entropy, one~notices in Figure~\ref{fig-opt-q-1}a that the smallest critical value of white noise occurs for $r\rightarrow 1$, which corresponds to the criteria based on Shannon entropy. Meanwhile, in Figure~\ref{fig-opt-q-1}b, the best criteria from Tsallis entropy are the ones for $q=2$ and $q=3$, which give an improvement to the Shannon-based criteria. Please note that within the interval $q\in [2;3]$ the line seems to be flat, meaning that any $q$ in this interval could be considered as an optimal value for the detection of steering from generalized entropies. However, this statement does not hold in general, as one can see in Figure~\ref{fig-opt-q-2}. It is true for the case of Werner states, whereas for the other considered states the optimal parameter values are $q=2$ and $q=3$ only. 

It is worth mentioning that the criteria for $q=2$ and $q=3$, in the case of $d=2$, are analytically the same. Also, for these values of $q$, they can be connected to the variance criteria from Refs.~\cite{Guhne2004b,Zhen2016}. To see this, consider an observable $A$ with eigenvalues $\pm 1$ and corresponding outcome probabilities $p_\pm$. The variance of the observable $A$ is given by
\be
\delta^2(A) = 1- \langle A\rangle^2 = 1 - (p_+ - p_-)^2 = 2 (1- p_+^2 - p_-^2) \sim S_2 (A).
\ee

The same relation can be found for $q=3$. This equivalence between variances and Tsallis entropies with $q=2$ and $q=3$ can be extended to the related steering criteria.

For the two-qubit Werner state, it is known that the optimal white noise threshold ($w_{crit}$) is $1/\sqrt{3}$ for three (orthogonal) projective measurements~\cite{Cavalcanti2009}. Interestingly, the criteria based on Tsallis entropy achieve these values with $q=2$ and $q=3$.

\begin{figure}[H]
\centering
\includegraphics[scale=0.55]{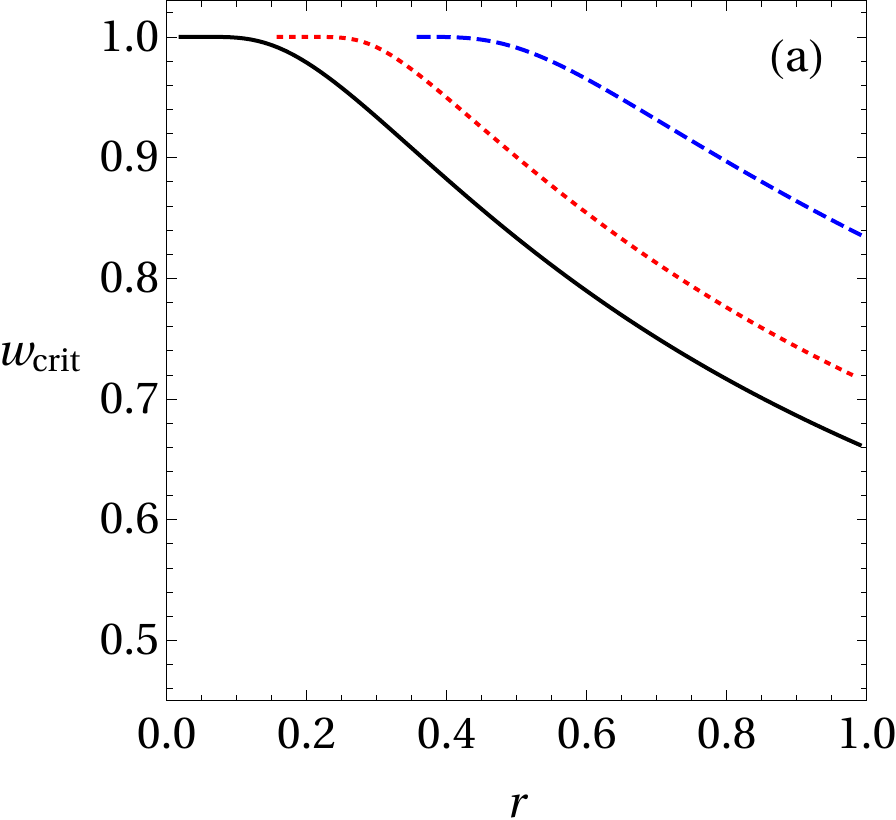}
\includegraphics[scale=0.55]{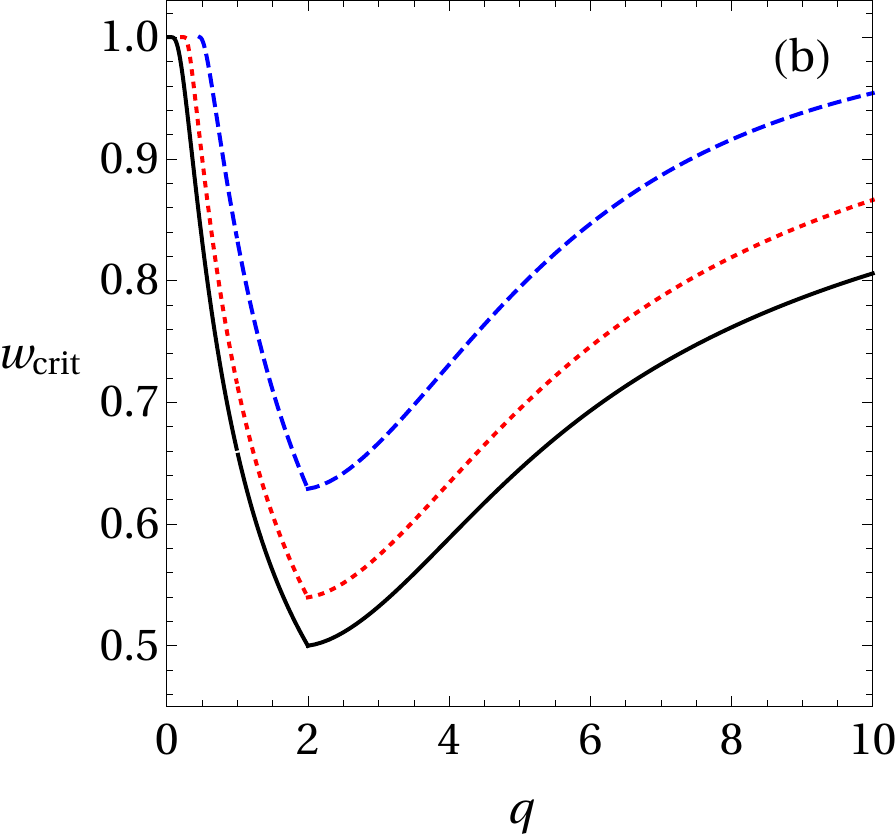}
\caption{The critical value $w$ for noisy two-qutrit entangled states $\varrho^{(3)}_{e_x} (w)$ for the detection of steering. The solid black line corresponds to the state with $x=1$, the dotted red line with $x=0.5$, and the dashed blue line with $x=0.2$, with (\textbf{a}) the criteria based on R\'enyi entropy~\eqref{tsc} and (\textbf{b}) on Tsallis entropy~\eqref{tsr}.}
\label{fig-op-q-qutrit}
\end{figure}

It is interesting to check whether the same optimal values of $r$ and $q$ also hold for some higher dimensional states. For this, consider noisy two-qutrit entangled states \mbox{$\varrho^{(3)}_{e_x} (w) = w \ket{\psi_x}\bra{\psi_x} + (1-w)\mathbbm{1}/9$}, where $\ket{\psi_x} = \frac{1}{\sqrt{2+x^2}}(\ket{00} + x\ket{11} + \ket{22})$. In Figure~\ref{fig-op-q-qutrit}, we analyse the states with $x= \{0.2, 0.5, 1\}$ when Alice performs a complete set of MUBs. One can see in Figure~\ref{fig-op-q-qutrit}a that for these states our criteria based on R\'enyi entropy are weaker than the ones based on Shannon entropy, similar to the case of two-qubit states. Interestingly, in Figure~\ref{fig-op-q-qutrit}b the optimal $q$ for the detection of steering using Tsallis entropy is only $q=2$ (and not $q=2$ {and} $q=3$ as in the two-qubit case).

We close this section with the conjecture that the criterion obtained from Tsallis entropy with $q=2$ is the best one to detect steerable states using the method proposed in this work for arbitrary (finite) dimensions. Our criteria based on R\'enyi entropy seems to be weaker than the one based on Shannon entropy (see Figures~\ref{fig-opt-q-1} and~\ref{fig-op-q-qutrit}) and, hence, we will not consider it further. Moreover, we will focus mainly on the criterion based on Tsallis entropy with $q=2$, but we also discuss results for different values of $q$.

\subsection{Isotropic States} 

The generalized entropic steering criteria are interesting for many scenarios, especially in the case of higher dimensional systems. Here, we address this scenario by applying our criteria to $d$-dimensional isotropic states~\cite{Horodecki1999}
\be\label{iso-states}
\varrho_{\rm iso} = 
\alpha\ket{\phi^+_d}\bra{\phi^+_d} + \frac{1-\alpha}{d^2}\mathbbm{1},
\ee
where $\ket{\phi^+} = ({1}/{\sqrt{d}})\sum_{i=0}^{d-1} \ket{i}\ket{i} $ is a maximally entangled state. These states are known to be entangled for $\alpha > 1/(d+1)$ and separable otherwise. To detect steering via our entropic criteria, we consider as measurements $m$ MUBs in dimension $d$ (provided that they exist). 

The marginal probabilities for this class of states are $p_i = 1/d$ for all $i$ and the joint probabilities are $p_{ii} = [1+(d-1)\alpha]/d^2$ (occurring $d$ times), and $p_{ij} = (1-\alpha)/d^2$ (for $i \neq j$ and occurring $d(d-1)$ times). Please note that since isotropic states are invariant under local unitary operators of the form $U\otimes U^\ast$ Ref.~\cite{Horodecki1999}, we choose Bob's measurements to be the conjugates of Alice's measurements. Inserting these probabilities in Equation~(\ref{tsc-prob}), the condition for non-steerability reads
\begin{equation}
\label{eq-iso}
\frac{m}{q-1}
\big\{1 - \frac{1}{d^q}[(1+(d-1)\alpha)^q + (d-1)(1-\alpha)^q]\big\} 
\geq \BB_B^{(q)},
\end{equation}
where $\BB_B^{(q)}$ is given in Equation~\eqref{boundq2} and~\eqref{bound-t-general} [in the limit of $q\rightarrow 1$, we use the bounds from Equation~\eqref{boundq1}]. One can see that Equation~\eqref{eq-iso} is valid for any dimension $d$, and depends only on the parameter $q$ and the number of MUBs $m$.

\begin{figure}[H]
\centering
\includegraphics[scale=0.6]{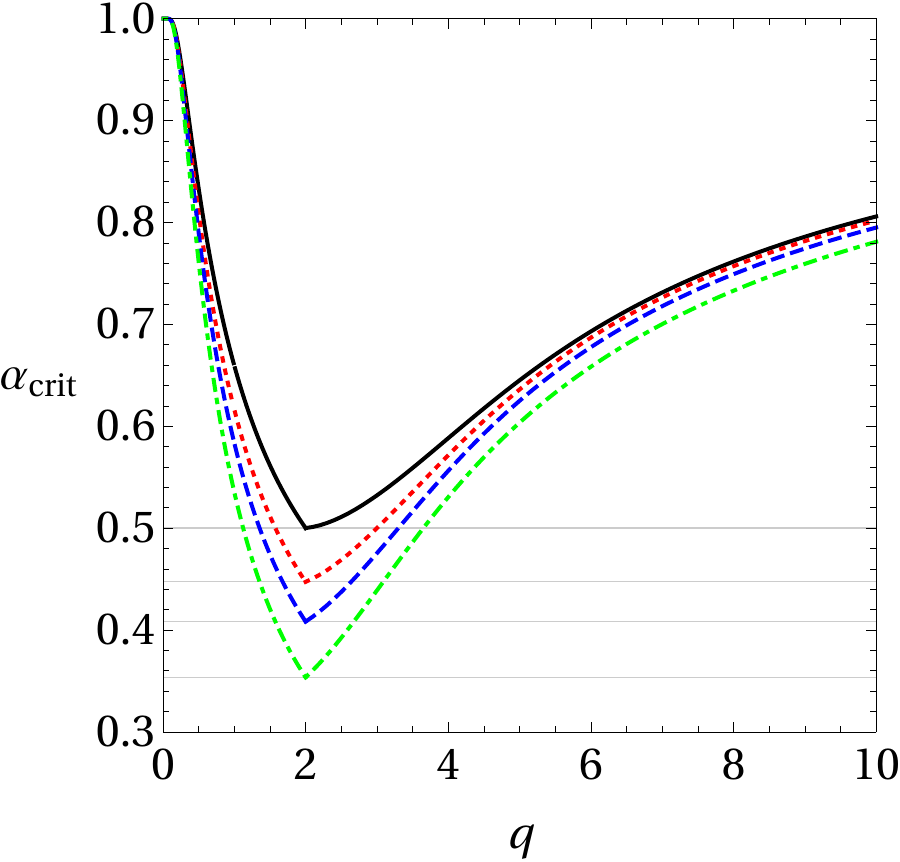}
\caption{The critical value of white noise $\alpha$ of states in Equation~\eqref{iso-states} as function of the Tsallis parameter $q$, considering a complete set of MUBs. Here, the solid black line corresponds to $d=3$, the dotted red line to $d=4$, the dashed blue line to $d=5$, and the dot-dashed green line to $d=7$. The optimal value for the detection of steerability is given by $q=2$.}
\label{fig-iso-states-q}
\end{figure}

Numerical investigations suggest that the criterion is strongest for $q=2$, as one can see in Figure~\ref{fig-iso-states-q}. For this value of $q$ the violation of Equation~(\ref{eq-iso}) occurs for $\alpha > 1/\sqrt{m}$. For a complete set of MUBs $(m = d+1)$ (with $d$ being a power of a prime) the violation happens for $\alpha > 1/\sqrt{d+1}$. For example, if we consider $d=2$ (qubits), isotropic states are equivalent to Werner states~\cite{Werner1989}. For a complete set of MUBs the violation of our criteria occurs for $\alpha > 1/\sqrt{3} \approx 0.577$, which is known to be the optimal threshold~\cite{Cavalcanti2009} for three MUBs.

\begin{figure}[H]
\centering
\includegraphics[scale=0.6]{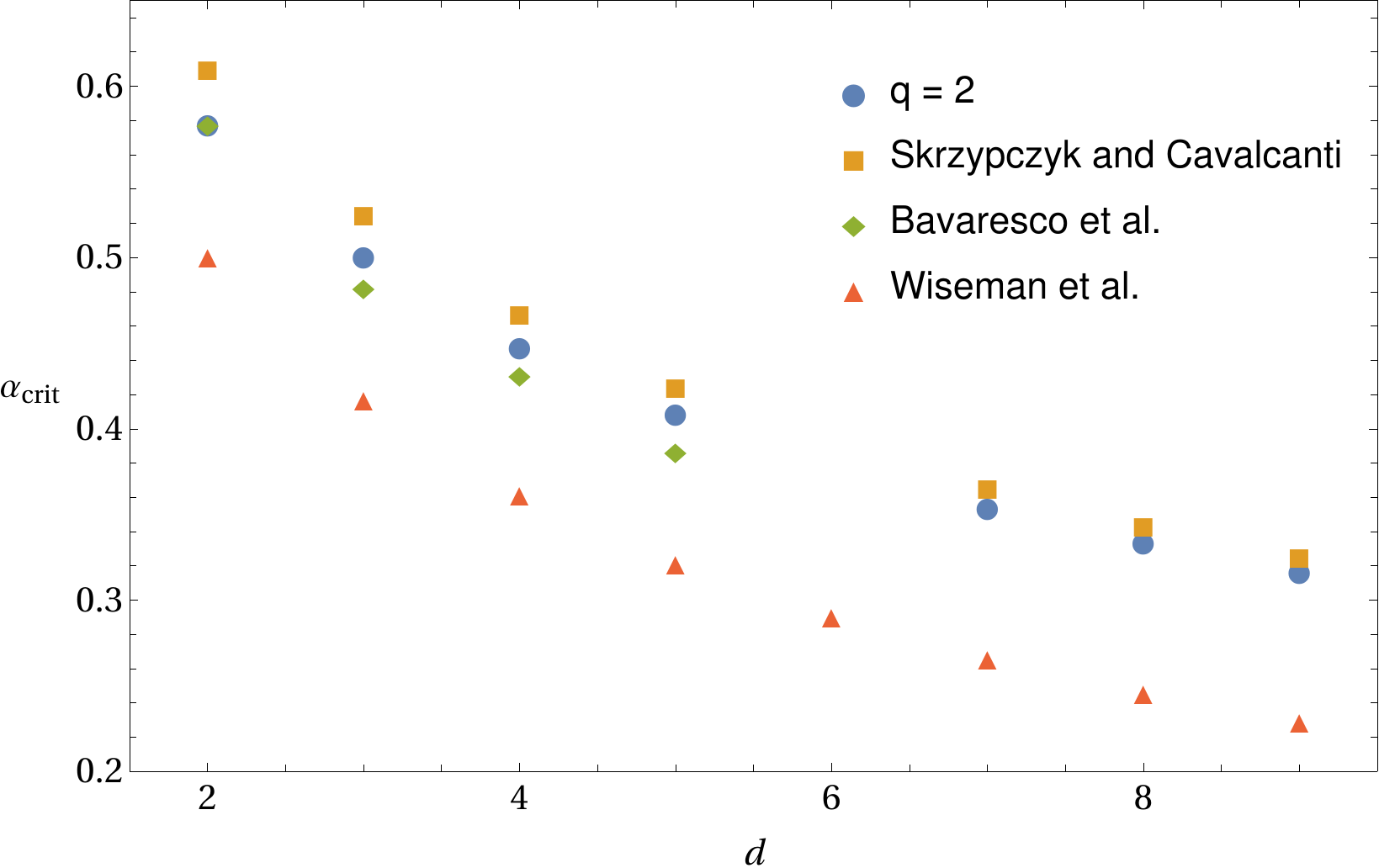}
\caption{The critical value of white noise $\alpha$ for 
different dimensions $d$, considering a complete set of MUBs. In this 
plot, blue circles correspond to our criterion in Equation~(\ref{eq-iso})
for $q=2$. The yellow squares 
correspond to the results for the inequality presented in Ref.~\cite{Cavalcanti2015} and the green diamonds in Ref.~\cite{Bavaresco2017}, where $\alpha_{\textrm{crit}}$ was calculated via SDP (numerical method). Below the red triangles the existence of an LHS model for all projective measurements (i.e. infinite amount of measurements instead of $d+1$ MUBs) is known~\cite{wiseman}. Please note that Ref.~\cite{wiseman} is given for comparison, this is not a steering criterion, but a bound on any criterion.}
\label{fig1}
\end{figure}

Now, we are able to compare our results with two others which investigated steering for the class of isotropic states and MUBs. In Ref.~\cite{Cavalcanti2015}, a steering inequality has been presented which is violated for $\alpha > (d^{3/2}-1)/(d^2 -1)$, whereas in Ref.~\cite{Bavaresco2017} the authors used semi-definite programming for this task. In Figure~\ref{fig1}, we show this comparison. Please note that we present only the results for $q=2$, which is the conjectured optimal value (Figure~\ref{fig-iso-states-q}). From Figure~\ref{fig1}, one sees that our criterion is stronger than the one from Ref.~\cite{Cavalcanti2015}. For $3\leq d \leq 5$ a better threshold than ours was obtained in Ref.~\cite{Bavaresco2017}, but it is worth mentioning that our criteria directly use probability distributions from a few measurements, without the need of performing full tomography on Bob's conditional state. In addition the numerical approach becomes computationally more demanding when increasing the number of variables.

\subsection{General Two-Qubit States}
Let us now consider the application of our method to general two-qubit 
states. Any two-qubit state can, after application of local unitaries, be
written as
\begin{equation}
\label{eq-rho}
\varrho_{AB} =
\frac{1}{4}
\big[\mathbbm{1}\otimes\mathbbm{1}+
(\vec{a}\vec{\sigma})\otimes\mathbbm{1}
+\mathbbm{1}\otimes(\vec{b}\vec{\sigma})
+\sum_{i=1}^3 c_i \sigma_i\otimes \sigma_i\big],
\end{equation}
where $\vec{a},\vec{b},\vec{c} \in\mathbb{R}^3$ are vectors with 
norm less than one, $\vec{\sigma}$ is a vector composed of the 
Pauli matrices and $(\vec{a}\vec{\sigma})= \sum_i a_i \sigma_i$. 
We assume that Alice performs projective measurements with 
effects \mbox{$P^A_m = [\mathbbm{1} + \mu_m (\vec{u}_m\vec{\sigma})]/2$} 
and Bob with effects 
$P_m^B = [\mathbbm{1} + \nu_m (\vec{v}_m\vec{\sigma})]/2$, where $\mu_m, \nu_m = \pm 1$
and $\{\vec{u},\vec{v}\}$ are unit vectors in $\mathbb{R}^3$. We have the following probabilities:
\begin{eqnarray}
p(\mu_m) &=& \textrm{Tr}[(P_m^A\otimes \mathbbm{1})\varrho_{AB}] = \frac{1}{2}(1 + \mu_m(\vec{a}\vec{u}_m)), \nonumber \\
p(\mu_m,\nu_m) &=& \textrm{Tr}[(P_m^A\otimes P_m^B)\varrho_{AB}] = \frac{1}{4}(1 + \mu_m(\vec{a}\vec{u}_m) + \nu_m(\vec{b}\vec{v}_m) + \mu_m\nu_m T_m),\nonumber
\end{eqnarray}
where $T_{m}=\sum_{i=1}^3 c_i u_{im}v_{im}$. Now Equation~\eqref{tsc-prob} can be written as
\be
\sum_m
\Big[
1 
-
\sum_{\mu_m, \nu_m}
\frac{[1 
+ \mu_m(\vec{a}\vec{u}_m) 
+ \nu_m(\vec{b}\vec{v}_m) 
+ \mu_m\nu_m T_{m}]^q}
{2^{q+1}[1+\mu_m(\vec{a}\vec{u}_m)]^{q-1}}
\Big]
\geq (q-1) \BB_B^{(q)}.
\ee

The optimization over measurements in this criterion for a general two-qubit state
is involving. We will focus on the simple case of Pauli measurements,
meaning that \mbox{$\vec{u}_m = \vec{v}_m = \{(1,0,0)^T, (0,1,0)^T,(0,0,1)^T\}$} 
and $q=2$. Then we have the following inequality
\begin{equation}
\label{criteria-general}
\sum_{i=1}^3\left[\frac{1 - a_i^2 - b_i^2 - c_i^2 + 2a_ib_ic_i}{2(1-a_i^2)}\right] \geq 1,
\end{equation}
the violation of which implies steerability.

Now, we can compare our criteria with other proposals for the detection 
of steerable two-qubit states using three measurements. 
The criterion from \cite{Guhne2004} (see Equation~\eqref{eq-criteria-sep}) proves steerability 
if $\sum_{i=1}^3 c_i^2 > 1$, and from the linear criteria~\cite{wiseman, Costa2016} steerability follows if $({\sum_{i=1}^3 c_i^2})^{1/2} > 1$.  
Not surprisingly, Equation~(\ref{criteria-general}) is stronger, 
since it uses more information about the state. The claim can be made
hard by analyzing $10^6$ (Hilbert-Schmidt) random two-qubit states~\cite{random-ensemble}. 
$94.34\%$ of the 
states do not violate any of the criteria, $3.81\%$ are steerable according 
to all criteria, $1.85\%$ violate only criterion~\eqref{criteria-general}, and~none of the states violates the linear criteria without violating~\eqref{criteria-general}.

A special case of two-qubit states are the Bell diagonal ones, which 
can be obtained if we set $\vec{a} = \vec{b} = 0$ 
in Equation~\eqref{eq-rho}. For this class of states it is easy to see that the 
three criteria are equivalent. Note, moreover, that a necessary 
and sufficient condition for steerability of this class of states with all projective measurements has recently been found~\cite{chau2016}.

\subsection{One-Way Steerable States}
As an example of weakly steerable states that can be detected with 
our methods we take one-way steerable states, i.e., states that 
are steerable from Alice to Bob but not the other way around. More~specifically, we 
consider the family of states given as
\begin{equation}
\label{one-way-state}
\varrho_{AB} = \beta\ket{\psi (\theta)}\bra{\psi (\theta)}+(1-\beta)\frac{\mathbbm{1}}{2}\otimes \varrho_B^\theta ,
\end{equation}
where $\ket{\psi (\theta)} = \cos (\theta)\ket{00} + \sin (\theta)\ket{11}$ 
and $\varrho_B^\theta = \text{Tr}_{A}[\ket{\psi (\theta)}\bra{\psi (\theta)}]$. 
It is known that states with $\theta \in [0,\pi/4]$ and 
$\cos^2 (2\theta) \geq ({2\beta-1})({(2-\beta)\beta^3})$ are not steerable from Bob to Alice considering all possible projective measurements~\cite{Bowles2016}, while Alice can steer 
Bob whenever $\beta > 1/2$. 

\begin{figure}[H]
\centering
\includegraphics[scale=0.65]{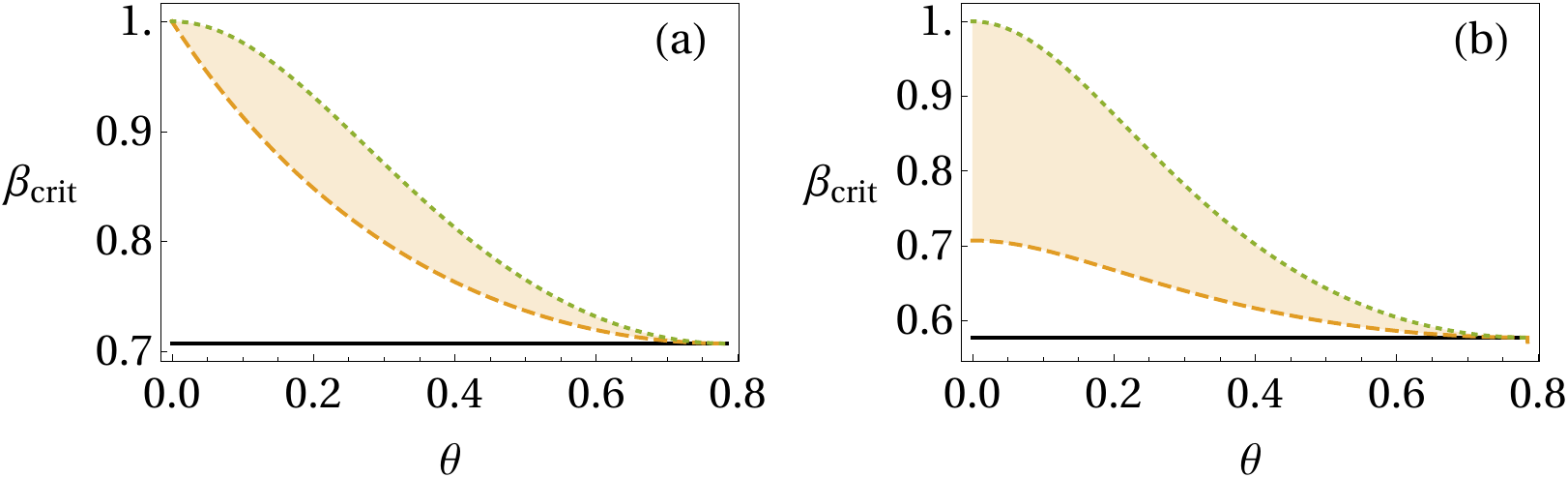}
\caption{One-way steerability of states~\eqref{one-way-state} for (\textbf{a}) two and (\textbf{b}) three measurement settings. The shaded area is the region where our criterion detects these weakly steerable states.}
\label{fig4}
\end{figure}

Considering two measurement settings, we have that this state is one-way steerable for \mbox{$1/\sqrt{2} < \beta \leq \beta_{\rm max}^{(2)}$ with $\beta_{\rm max}^{(2)} = [{1+\sin^2(2\theta)}]^{-1/2}$}, and for three measurement settings, this 
state is one way-steerable for 
${1}/{\sqrt{3}} < \beta \leq \beta_{\rm max}^{(3)}$
with $\beta_{\rm max}^{(3)} = [{1+2\sin^2(2\theta)}]^{-1/2}$~\cite{Xiao2017}. 
For our entropic steering criterion~\eqref{tsc} with $q=2$ we find that this state is one-way steerable in the range
\be
\label{esc-one-way-2}
\frac{\sqrt{1+\tan^2 (\theta)}}{1+\tan(\theta)}< 
\beta 
\leq \beta_{\rm max}^{(2)},
\ee
for two Pauli measurements ($\sigma_x$, $\sigma_z$), and
\be
\label{esc-one-way-3}
\frac{1}{2\cos(2\theta)}\sqrt{3-\sqrt{1+8\sin^2 (2\theta)}}
< 
\beta 
\leq \beta_{\rm max}^{(3)},
\ee
for three Pauli measurements ($\sigma_x,\sigma_y,\sigma_z$).
For any $\theta$ this gives a non-empty interval of $\beta$ for which our criterion 
detects these
weakly steerable states. In Figure~\ref{fig4}, we show the range of one-way steerability considering two and three measurement settings.

\subsection{Bound Entangled States}

It is also interesting to investigate whether the entropic steering criteria from generalized entropies are able to detect steerability of bound entangled states, which is related to the stronger version of Peres conjecture~\cite{Peres1999,pusey,PS2014}. The conjecture states the possibility of constructing local models for bound entangled states and it was proven wrong in Refs.~\cite{moroder,Vertesi2014}.

For this task, we investigated the following class of states presented in~\cite{moroder}
\be
\varrho_{BES} = \lambda_1 \ket{\psi_1}\bra{\psi_1} + \lambda_2 \ket{\psi_2}\bra{\psi_2} + \lambda_3 (\ket{\psi_3}\bra{\psi_3} + \ket{\tilde{\psi}_3}\bra{\tilde{\psi}_3}),
\ee
with the following normalized states
\begin{eqnarray}
\ket{\psi_1} &=& (\ket{12}+\ket{21})/\sqrt{2}, \nonumber \\
\ket{\psi_2} &=& (\ket{00}+\ket{11}-\ket{22})/\sqrt{3}, \nonumber \\
\ket{\psi_3} &=& m_1\ket{01} + m_2\ket{10} + m_3(\ket{11} + \ket{22}), \nonumber \\
\ket{\tilde{\psi}_3} &=& m_1\ket{02} - m_2\ket{20} + m_3(\ket{21} - \ket{12}),
\end{eqnarray}
where $m_{1(2)}\geq 0$ and $m_3 = \sqrt{(1-m_1^2-m_2^2)/2}$. This class of states has a positive partial transpose if the eigenvalues are fixed as
\begin{eqnarray}
\lambda_1 &=& 1 - (2+3m_1 m_2)/N, \nonumber \\
\lambda_2 &=& 3m_1m_2/N, \nonumber \\
\lambda_3 &=& 1/N,
\end{eqnarray}
with $N = 4 - 2m_1^2 + m_1 m_2 - 2m_2^2$ and $m_1^2 + m_2^2 + m_1m_2 \leq 1$. In Ref.~\cite{moroder}, the authors show that this class of states is steerable for certain measurements. Now, to check whether the generalized entropic criteria are also able to detect the steerability of such states, consider that we perform the following two MUBs on Alice's and Bob's system~\cite{moroder}:
\begin{eqnarray}\label{meas1}
M_1^1 &=& [1/\sqrt{3},-1/\sqrt{6},-1/\sqrt{2}], \nonumber \\
M_2^1 &=& [1/\sqrt{3},-1/\sqrt{6},1/\sqrt{2}], \nonumber \\
M_3^1 &=& [1/\sqrt{3},\sqrt{2/3}],
\end{eqnarray}
for measurement $m=1$, and
\begin{eqnarray}\label{meas2}
M_1^2 &=& [1,0,0], \nonumber \\
M_2^2 &=& [0, q/\sqrt{2},i q/\sqrt{2}], \nonumber \\
M_3^2 &=& [0, q^*/\sqrt{2},-iq^*/\sqrt{2}],
\end{eqnarray}
for measurement $m=2$. These rotated MUBs are given by the symmetry of the above class of states. Since they are MUBs, the bound $\CC^{(2)} = 1$ holds. 

\begin{figure}[H]
\centering
\includegraphics[scale=0.6]{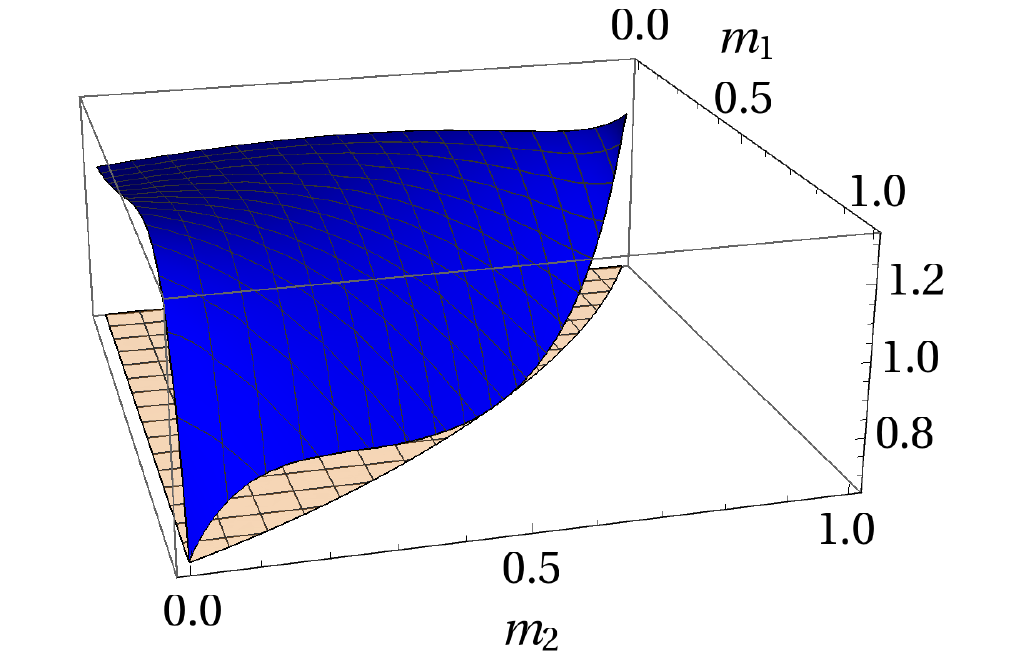}
\caption{Plot of Equation~\eqref{tsc} in terms of $m_1$ and $m_2$ with $q=2$ (blue curve) for $\varrho_{BES}$ in the region $m_1^2 + m_2^2 + m_1m_2 \leq 1$. The opaque flat plot is the entropic uncertainty bound for $q=2$ and the measurements given by Equations~\eqref{meas1} and~\eqref{meas2}. From the plot one can see that there is no violation of Equation~\eqref{tsc} for any state in this family.}
\label{fig-bound}
\end{figure}

In Figure~\ref{fig-bound}, one can see that no violation for this specific class of bound entangled states occurs (given the above measurements). Surprisingly, performing more measurements makes no difference for the detection of steerability using our entropic steering criterion. This situation can be explained by the symmetry of such states, i.e., with the addition of more mutually unbiased measurements the entropic uncertainty bound increases with the same rate as the l.h.s. of criterion~\eqref{tsc}. The same result can also be obtained by a numerical optimization over random unitaries applied in the standard MUBs, where we use the parametrization given in Ref.~\cite{Bronzan1988}. In this sense, it remains as an open question if the criterion is able to detect steerable bound entangled states.

\section{Multipartite Scenario}
\label{sec-multi}

In this section we extend the generalized entropic criteria to the case of tripartite systems. For~such systems, one can consider two different steering scenarios: either Alice tries to steer Bob and Charlie or Alice and Bob try to steer Charlie. In the latter scenario, one should notice that there is a difference regarding the kind of measurements Alice and Bob perform: local or global ones. In this section we consider all these cases and derive generalized multipartite steering criteria from the Tsallis entropy. This specific choice of entropy is given by the examples presented in the previous sections, where the criteria based on R\'enyi entropy were found weak in comparison to the one based on Shannon entropy, which, by extension, is included in the Tsallis entropy.

A proposal for multipartite steering using EURs based on Shannon entropy has been recently introduced in Ref.~\cite{Riccardi2017}. Here, we derive our criteria from a different perspective considering a general approach via Tsallis entropy.

\subsection{Steering from Alice to Bob and Charlie}

Let us first focus on the scenario where Alice tries to steer Bob and Charlie. Consider the quantity 
\be
F_q(A,B,C) = - D_q(A\otimes B\otimes C||A\otimes \mathbbm{I}_B\otimes \mathbbm{I}_C),
\ee
where $\mathbbm{I}_{B(C)}$ are equal distributions with $p_j = 1/N_B$ and $p_k = 1/N_C$, respectively. Writing this in terms of probabilities gives [see also Equations~\eqref{tsallis-1} and \eqref{tsallis-2}]
\begin{eqnarray}
F_q (A,B,C) &=& \sum_{i,j,k}p_{ijk}\ln_q \left(\frac{p_i/N_{BC}}{p_{ijk}}\right) \nonumber \\
&=& \frac{x_{BC}}{1-q}+(1+x_{BC})[S_q(A,B,C)-S_q(A) + (1-q)T_q^{(1)}(A,B,C)],
\end{eqnarray}
where $N_{BC} = N_B N_C$, $x_{BC} = (N_B N_C)^{q-1} - 1$ and
\be
T_q^{(1)}(A,B,C) = \sum_i p_i^q (\ln_q(p_i))^2 - \sum_{i,j,k}p_{ijk}^q \ln_q(p_i)\ln_q(p_{ijk}),
\ee
is the correction term.

Now, from the LHS model~\eqref{lhs-model-1-2}, the probability distribution $p(i|A,\lambda){ p_Q}(j|B,\lambda){ p_Q}(k|C,\lambda)$ with a fixed $\lambda$ yields
\be
F_q^{(\lambda)} (A,B,C) = \frac{x_{BC}}{1-q}+(1+x_{BC})S_q^{(\lambda)} (B,C),
\ee
with $S_q^{(\lambda)} (B,C) = S_q^{(\lambda)} + S_q^{(\lambda)} (C) + (1-q)S_q^{(\lambda)} (B)S_q^{(\lambda)} (C)$. For a given set of measurements $B_m \otimes C_m$ one has an EUR
\be
\sum_m S_q^{(\lambda)} (B_m,C_m) \geq \CC_{BC}^{(q)},
\ee
where $\CC_{BC}^{(q)}$ is some entropic uncertainty bound for the observables $B_m \otimes C_m$. Since $S_q$ is a concave function, the same bound holds for convex combinations of product distributions $p(i|A,\lambda){ p_Q}(j|B,\lambda){ p_Q}(k|C,\lambda)$. Connecting the above results, the generalized multipartite steering criteria from Alice to Bob and Charlie are given by
\be\label{eq-crit-1-2}
\sum_m [S_q(B_m,C_m|A_m) + (1-q)T_q^{(1)}(A_m,B_m,C_m)] \geq \CC_{BC}^{(q)},
\ee
where $S_q(B_m,C_m|A_m) = S_q(B_m,C_m) - S_q(A_m)$ is the conditional Tsallis entropy. In terms of probabilities, these criteria can be written as
\be
\frac{1}{q-1}\left[\sum_m \left(1-\sum_{i,j,k}\frac{(p_{ijk}^{(m)})^q}{(p_{i}^{(m)})^{q-1}}\right)\right] \geq \CC_{BC}^{(q)}.
\ee

Note here that we define tripartite steering from Alice to Bob and Charlie from the LHS model given in Equation~\eqref{lhs-model-1-2}, and in this case we should consider the EUR bounds for separable states, see~for example Equations~\eqref{bound-s-sep}, \eqref{bound-s-3-sep},\eqref{bound-sep-general}, \eqref{bound-q-2} and \eqref{bound-q-3}, for the case of qubits and Pauli measurements. Moreover, if we consider the bound where we allow the state of Bob and Charlie to be entangled, which leads effectively to the scenario of bipartite steering, the bound for three measurement settings changes for Shannon entropy (see Equation~\eqref{bound-s-3-ent}). For Tsallis entropy, the EUR bound differs by non-separable states in the range of $1\leq q < 2$, for three measurement settings (see Figure~\ref{fig-bound-sep-ent}). These different scenarios will be discussed further for some class of states in the next section.

\subsection{Steering from Alice and Bob to Charlie}

Let us now consider the scenario where Alice and Bob try to steer Charlie. Here, we follow the definition of tripartite steering given through Equation~(\ref{lhs-model-2-1-local}). To start with, consider the quantity 
\be
F_q(A,B,C) = - D_q(A\otimes B\otimes C||A\otimes B \otimes \mathbbm{I}_C),
\ee
where $\mathbbm{I}_{C}$ represents a uniform distribution with $p_k = 1/N_C$. In terms of probabilities one gets
\begin{eqnarray}
F_q (A,B,C) &=& \sum_{i,j,k}p_{ijk}\ln_q \left(\frac{p_{ij}/N_C}{p_{ijk}}\right) \nonumber \\
&=& \frac{x_C}{1-q}+(1+x_C)[S_q(A,B,C)-S_q(A,B) + (1-q)T_q^{(2)}(A,B,C)],
\end{eqnarray}
where $x_C = N_C^{q-1} - 1$ and
\be
T_q^{(2)}(A,B,C) = \sum_{i,j} p_{ij}^q (\ln_q(p_{ij}))^2 - \sum_{i,j,k}p_{ijk}^q \ln_q(p_{ij})\ln_q(p_{ijk}),
\ee
is the correction term.

Assuming that one has the LHS model from Equation~\eqref{lhs-model-2-1-local} and considering the probability distribution $p(i|A,\lambda)p(j|B,\lambda){ p_Q}(k|C,\lambda)$ with a fixed $\lambda$ one gets
\be
F_q^{(\lambda)} (A,B,C) = \frac{x_{C}}{1-q}+(1+x_{C})S_q^{(\lambda)} (C).
\ee

For a given set of measurements $\{C_m\}_m$ one has an EUR
\be
\sum_m S_q^{(\lambda)} (C_m) \geq \BB_{C}^{(q)},
\ee
where $\BB_{C}^{(q)}$ is some entropic bound for the observables $\{C_m\}_m$. Since $S_q$ is concave function, the~same bound holds for convex combinations of product distributions $p(i|A,\lambda)p(j|B,\lambda){ p_Q}(k|C,\lambda)$. Connecting the above results, the generalized multipartite steering criteria from Alice to Bob and Charlie are given by
\be\label{eq-crit-2-1-local}
 \sum_m [S_q(A_m,B_m,C_m)-S_q(A_m,B_m) + (1-q)T_q^{(2)}(A_m,B_m,C_m)] \geq \BB_{C}^{(q)}.
\ee

In terms of probabilities, these criteria can be written as
\be
\frac{1}{q-1}\left[\sum_m \left(1-\sum_{i,j,k}\frac{(p_{ijk}^{(m)})^q}{(p_{ij}^{(m)})^{q-1}}\right)\right] \geq \BB_{C}^{(q)}.
\ee

In this scenario, this framework is not able to distinguish between bipartite and tripartite steering. This comes from the fact that if we consider the LHS model given in~\eqref{lhs-model-2-1-global}, with product distributions $p(i,j|A,B,\lambda){ p_Q}(k|C,\lambda)$, we obtain the same criteria.

\subsection{Applications}

For the application of the multipartite entropic steering criteria, we consider systems of three qubits with Pauli measurements. We focus our discussion on GHZ and W states. A noisy GHZ state is defined as
\begin{equation}
\rho_{GHZ} = \gamma \ket{GHZ}\bra{GHZ} + \frac{1-\gamma}{8}\mathbbm{1},
\end{equation}
\textls[-15]{where $\ket{GHZ} = \frac{1}{\sqrt{2}}(\ket{000}+\ket{111})$. This state is known to be not fully separable iff $\gamma > 1/5$~\cite{Schack2000,Cirac2000} and to be Bell nonlocal for $\gamma > 1/2$ for two and three measurements per site~\cite{Gruca2010}. A noisy W state reads}
\be
\rho_W = \delta \ket{W}\bra{W} + \frac{1-\delta}{8}\mathbbm{1},
\ee
where $\ket{W} = \frac{1}{\sqrt{3}} (\ket{100} + \ket{010} + \ket{001})$, being entangled for ${\delta} > \sqrt{3}/(8+\sqrt{3}) \approx 0.178$ and fully separable for ${\delta}\leq 0.177$~\cite{Chen2012}. This state is Bell nonlocal for $\delta > 0.6442$ for two measurements per site and $\delta > 0.6048$ for three measurements~\cite{Gruca2010}. Here, we are interested in the critical amount of white noise for the violation of criteria \eqref{eq-crit-1-2} and \eqref{eq-crit-2-1-local} with the aforementioned of measurements (together with an optimization over local unitaries).

Let us start discussing the results for the scenario of steering from Alice to Bob and Charlie. As~mentioned above, we can distinguish the results into two different steering scenarios-bipartite and tripartite-depending on the considered LHS model and, consequently, the associated entropic bounds. 

For the case of noisy GHZ states we have the following results for two measurement settings. Considering that Bob and Charlie always perform the same measurements (restriction given by the EUR bounds), violation of the criterion~\eqref{eq-crit-1-2} is found for 
\begin{align}\label{eq-ghz-two-meas}
A_1 = B_1 = C_1 = \sigma_x, \nonumber \\
A_2 = B_2 = C_2 = \sigma_z, 
\end{align}
with $\gamma > \gamma_{crit}^{(1)} \approx 0.8631$ and $\gamma > \gamma_{crit}^{(2)} \approx 0.866$, where the notation $\gamma^{(q)}$ is used to distinguish between Shannon and Tsallis entropies. For three measurement settings, we choose the measurements as
\begin{align}\label{eq-ghz-three-meas}
A_1 = A_2 = B_1 = C_1 = \sigma_x, \nonumber \\
B_2 = C_2 = \sigma_y, \\
A_3 = B_3 = C_3 = \sigma_z, \nonumber
\end{align}
and the state is steerable from Alice to Bob and Charlie for $\gamma > \gamma_{crit}^{(1)} \approx 0.7642$ (for the bound~\eqref{bound-s-3-sep}) and $\gamma > \gamma_{crit}^{(1)} \approx 0.909$ (for the bound~\eqref{bound-q-3} with $q\rightarrow 1$). Using the criteria from Tsallis entropy (for the bound~\eqref{bound-q-3}) $\gamma > \gamma_{crit}^{(2)} \approx 0.775$. Please note that the best noise threshold is obtained using Shannon entropy and the bound for separable states, which leads to a ``truly'' tripartite steering scenario. In other words, the criteria obtained from Shannon entropy is sensitive to this distinction and demonstrates that is ``easier'' for Alice to steer Bob and Charlie if they share a separable state. In contrast, the criteria from Tsallis entropy with $q=2$ is not sensitive (indifferent) within these different scenarios.

For the noisy W states, we have the following results for two-measurement settings. The optimal measurements are the ones given by Equation~\eqref{eq-ghz-two-meas}. Violation of the criteria occurs for $\delta > \delta_{crit}^{(1)} \approx 0.9814$, and no violation was found for $q=2$. Considering three-measurement setting, the optimal set of measurements is 
\begin{align}\label{eq-w-three-meas}
A_1 = B_1 = C_1 = \sigma_x, \nonumber \\
A_2 = B_2 = C_2 = \sigma_y, \\
A_3 = B_3 = C_3 = \sigma_z, \nonumber
\end{align}
and there is no violation for the criteria with the bound~\eqref{bound-s-3-ent}, but $\delta >\delta_{crit}^{(1)} \approx 0.8523$ for the criteria with the bound~\eqref{bound-s-3-sep}, and $\delta > \delta_{crit}^{(2)} \approx 0.8366$ for the bound~\eqref{bound-q-3}. The best threshold for steerability occurs for the criterion based on Tsallis entropy (contrary to the results found for noisy GHZ states), although the criterion does not distinguish between bipartite and tripartite LHS models.

Now, consider steering from Alice and Bob to Charlie. As mentioned above, in this scenario we have no distinction between bipartite and tripartite steering, since both models lead to the same criteria. However, it is possible to explore the difference between performing local and global measurements.

Let us first discuss the results for local measurements. For noisy GHZ states and two measurement settings we use the measurements from Equation~\eqref{eq-ghz-two-meas}. Steerability from Alice and Bob to Charlie occurs for $\gamma > \gamma_{crit}^{(1)} \approx 0.7476$ and $\gamma > \gamma_{crit}^{(2)} \approx 0.6751$. For three measurement settings, considering the measurements from Equation~\eqref{eq-ghz-three-meas}, one has $\gamma >  \gamma_{crit}^{(1)} \approx 0.6247$ and $\gamma > \gamma_{crit}^{(2)} \approx 0.5514$. 

For noisy W states we use the measurements
\begin{align}\label{eq-w-two-meas}
A_1 = C_1 = \sigma_x, \nonumber \\
A_2 = B_1 = B_2 = C_2 = \sigma_z. 
\end{align}
Here steering occurs for $\delta > \delta_{crit}^{(1)} \approx 0.818$ and $\delta >  \delta_{crit}^{(2)} \approx 0.75$. For three measurement settings we take
\begin{align}\label{eq-w-three-meas-2}
A_1 = C_1 = \sigma_x, \nonumber \\
A_2 = C_2 = \sigma_y,  \nonumber \\
A_3 = B_1 = B_2 = B_3 = C_3 = \sigma_z.
\end{align}

The corresponding thresholds are $\delta_{crit}^{(1)} \approx 0.698$ and $\delta_{crit}^{(2)} \approx 0.623$.  In this scenario, one can notice that increasing the number of measurements and choosing $q=2$, one is able to detect more steering for both families of states-in the same way as in bipartite steering. 

Now, let us explore the scenario where Alice and Bob perform global measurements. For this, we~consider MUBs in dimension 4 for the global measurements and Pauli measurements to be performed in Charlie system. A possible set of MUBs in dimension 4 is given by
\begingroup\makeatletter\def\f@size{8}\check@mathfonts
\def\maketag@@@#1{\hbox{\m@th\fontsize{10}{10}\selectfont\normalfont#1}}%

\begin{eqnarray}
M_1 &=& \left(
\begin{array}{cccc}
1 & 0 & 0 & 0 \\
0 & 1 & 0 & 0 \\
0 & 0 & 1 & 0 \\
0 & 0 & 0 & 1
\end{array}
\right), \quad
M_2 = \frac{1}{2}\left(
\begin{array}{cccc}
1 & 1 & 1 & 1 \\
1 & 1 & -1 & -1 \\
1 & -1 & -1 & 1 \\
1 & -1 & 1 & -1
\end{array}
\right), \quad
M_3 = \frac{1}{2}\left(
\begin{array}{cccc}
1 & 1 & 1 & 1 \\
-1 & -1 & 1 & 1 \\
-i & i & i & -i \\
-i & i & -i & i
\end{array}
\right), \nonumber \\
M_4 &=& \frac{1}{2}\left(
\begin{array}{cccc}
1 & 1 & 1 & 1 \\
-i & -i & i & i \\
-i & i & i & -i \\
-1 & 1 & -1 & 1
\end{array}
\right), \quad
M_5 = \frac{1}{2}\left(
\begin{array}{cccc}
1 & 1 & 1 & 1 \\
-i & -i & i & i \\
-1 & 1 & -1 & 1 \\
-i & i & i & -i
\end{array}
\right).
\end{eqnarray}
\endgroup

From this set, we can choose within five measurements, while for the measurements of Charlie's system we can choose within three Pauli measurements. The task is to find the optimal combination which shows the best threshold for steerability in this scenario.

Considering the noisy GHZ states, the optimal two-measurement choice (from the given set) is $(AB)_1 = M_1$, $(AB)_2 = M_2$, and $C_1 = \sigma_z$, $C_2 = \sigma_x$, and the optimal three-measurement setting is the same as the two-measurement setting, with the addition of the third measurement $(AB)_3 = M_3$ and $C_3=\sigma_y$, which gives the same noise threshold found in the scenario of local measurements. Hence, the criterion is not able to detect a difference between local and global measurements for this specific family of states and set of measurements. 

However, this is not the case for the noisy W states. The optimal two-measurement setting (from the given set) is $(AB)_1 = M_1$, $(AB)_2 = M_2$, and $C_1 = \sigma_z$, $C_2 = \sigma_x$, with the noise threshold \mbox{$\delta_{crit}^{(1)} \approx 0.8571$} and $\delta_{crit}^{(2)} \approx 0.7802$. The optimal three-measurement setting is the same as the two-measurement setting, with the addition of the third measurement $(AB)_3 = M_4$ and $C_3=\sigma_y$, with~the noise threshold $\delta_{crit}^{(1)} \approx 0.7414$ and $\delta_{crit}^{(2)} \approx 0.6548$. These results show that for noisy W states, local measurements are able to detect steerability with smaller noise threshold while compared to global ones. This result shows that the standard MUBs are not a good choice of global measurements, since they should reveal steerability with lower thresholds while compared to local ones. 

Now, we are able to compare our results to the literature. For example, in the case of Shannon entropy and two measurement settings, we obtain the same results as the ones presented in Ref.~\cite{Riccardi2017} for noisy GHZ and W states and scenarios of steering from Alice to Bob and Charlie and Alice and Bob to Charlie. In the latter case, we were able to find a smaller threshold considering Tsallis entropy and $q=2$. However, if we compare our results with the ones in Ref.~\cite{Cavalcanti2015a}, our noise thresholds are bigger for all scenarios, and the same happens if we compare them with the nonlocality thresholds presented in Ref.~\cite{Gruca2010}.


\section{Conclusions} 

In this work we have extended to several directions the straightforward technique for the construction of strong steering criteria from EURs~\cite{Costa2018}. These criteria are easy to implement using a finite  set of measurement settings only, and do not need the use of semi-definite programming and full tomography on Bob's conditional states. We also show that they can be extended to multipartite systems, where different steering scenarios can be identified and evaluated. 

For future work, several directions seem promising. First, considering EURs in the presence of quantum memory~\cite{Berta2010} might improve the criteria. Second, connecting our results to measurement uncertainty relations for discrete observables~\cite{Toigo2018}. Third, making quantitative statements about steerability from steering criteria. Recently, some attempts in this direction have been pursued~\cite{Schnee2017}.


\vspace{6pt}

\authorcontributions{All authors contributed equally to this work.}

\funding{This work was supported by the DFG, the ERC (Consolidator Grant No. 683107/TempoQ) and the Finnish Cultural Foundation.}

\acknowledgments{We thank Chau Nguyen for the discussions about global versus local measurements, and for pointing us out the example of super-activation of steering. We also thank to Yichen Huang for bringing our attention to Ref.~\cite{Huang2010}, and Alberto Riccardi and René Schwonnek for discussions. We are also thankful for the comments on the earlier version of the manuscript by C. Jebaratnam.}

\conflictsofinterest{The authors declare no conflict of interest.} 

\reftitle{References}

\end{document}